\documentclass[12pt,onecolumn,twoside]{IEEEtran}
\usepackage{listings}
\usepackage{amsmath}
\usepackage{graphicx}
\usepackage{amsfonts}
\usepackage{array}
\usepackage{amssymb}
\usepackage{helvet}
\usepackage{color}
\usepackage{rawfonts,amsfonts,psfrag,flushend,geometry}
\usepackage{bm}
\usepackage{float}
\usepackage{cite}
\usepackage{multirow}
\usepackage{slashbox}

\begin{document}
\title{ Robust Beamforming for Physical Layer Security in BDMA Massive MIMO}
\author{Fengchao Zhu, Feifei Gao, Hai Lin,   Shi Jin, and Junhui Zhao
\thanks{F. Zhu is with High-Tech Institute of Xi'an, Xi'an, Shaanxi 710025, China (e-mail: fengchao\_zhu@126.com). F. Gao   is with the State Key Laboratory of Intelligent Technology and Systems,
Tsinghua National Laboratory for Information Science and Technology,
Department of Automation, Tsinghua University, Beijing, 100084, China
(e-mail: feifeigao@ieee.org). H. Lin is with the Department of Electrical and Information Systems, Osaka Prefecture University, Sakai, Osaka, Japan  (e-mail: hai.lin@ieee.org). S. Jin is with the National Communications Research Laboratory, Southeast University, Nanjing 210096, P. R. China (email: jinshi@seu.edu.cn). J. Zhao is with School of Information Engineering, East China Jiaotong University, Nanchang, 330013, China, and is also with School of Electronic and Information Engineering, Beijing Jiaotong University, Beijing, 100044, China (email: eeejhzhao@163.com).}
}
\maketitle

\vspace{-10mm}

\begin{abstract}
In this paper, we design robust beamforming to guarantee the physical layer security for a multiuser beam division multiple access (BDMA) massive multiple-input multiple-output (MIMO) system, when the channel estimation errors are taken into consideration.
With the aid of  artificial noise (AN), the proposed design are formulated as minimizing the transmit power of  the base station (BS), while providing  legal users and the eavesdropper (Eve) with different  signal-to-interference-plus-noise ratio (SINR).
It is strictly proved that, under  BDMA massive MIMO scheme, the initial non-convex optimization can be equivalently converted to a convex semi-definite programming (SDP) problem and the  optimal rank-one beamforming solutions can be guaranteed.
In stead of directly resorting to the convex tool, we make one step further by deriving the optimal beamforming direction and the optimal beamforming power allocation in closed-form, which greatly reduces the computational complexity and makes the proposed design practical for real world applications.
Simulation results are then provided to verify the efficiency of the proposed algorithm.
\end{abstract}

\newpage

\section{Introduction}\label{Sec.1}

Massive multiple-input multiple-output (MIMO) \cite{Marzetta2010} is one of the key technologies for 5G wireless communications \cite{Andrews2014}, which utilizes a
very large number of antennas at the base station (BS) to provide low power consumption, high spectral efficiency, security, and reliable linkage.
It was shown in \cite{Larsson2014} that the propagation channel vectors of different users become asymptotically orthogonal,
where the effects of uncorrelated noise and intra-cell interference will be eliminated when the number of antennas at the BS grows to infinity.
As a result, simple linear signal processing approaches, i.e., matched filter (MF) and zero-forcing beamforming can be used in massive MIMO systems \cite{Lu2014}.
However, a prerequisite of enjoying above benefits  is the availability of channel state information (CSI). There are three main approaches  of low-complex channel estimation algorithms
that referred to different expansion basis to represent the sparsity inside channels \cite{Xie20161,Xie20162,Xie20163,Yin2013,Adhikary2013,BDMA1,BDMA2}.
For example, the authors in \cite{Yin2013,Adhikary2013} applied a low-rank approximation of the channel covariance matrix to reduce the effective channel parameters of massive MIMO. The authors in \cite{Xie20161,Xie20162,Xie20163} proposed an angle division multiple access (ADMA) model and the massive MIMO channels could be represented  by a few channel gain and angular parameters.
Moreover, \cite{BDMA1,BDMA2} designed  a beam division multiple access (BDMA) scheme and a few orthogonal basis from discrete Fourier transform (DFT) are used to approximate the channel vectors.
Nevertheless,  perfect CSI are not available in practice due to: (i) all \cite{Yin2013,Adhikary2013,Xie20161,Xie20162,Xie20163,BDMA1,BDMA2} are based on approximately-sparse model and there
must be residue errors; (ii) additive noise always exist
 at the receiver. Therefore, the channel estimation errors should be taken into consideration in the subsequent transmission design.

On the other side, secrecy and privacy are also critical concerns for 5G communication systems,
where the physical layer security has drawn considerable interest since it can prevent eavesdropping
without upper layer data encryption. The information-theoretic approach to guarantee secrecy was
initiated by Wyner  \cite{Wynera1975}, where it was shown that confidential messages transmission could be achieved by exploiting the physical characteristics of the
wireless channel. Then the results of \cite{Wynera1975} was generalized to different channel models, i.e., the broadcast channels
\cite{Csiszar1978}, the single-input single-output (SISO) fading
channels \cite{Liang2008}, the multiple access channels (MAC)
\cite{Liang2006}, and the multiple-input multiple-output (MIMO)
channels \cite{Oggier2011}.
The authors of \cite{Zhu2014} first investigated the limiting physical layer security performance when the number of antennas approaches infinity in massive MIMO systems, and then \cite{Chen2015,Wang2015} studied the physical layer security in massive MIMO relay channel and massive MIMO Rician Channel, respectively.
The authors of \cite{Wanghm2016,Zhu12016,Wu2016} considered the asymptotic achievable secrecy rate in massive MIMO systems,
where it was shown that under certain orthogonality conditions,
the secrecy rate loss introduced by the eavesdropper could be completely mitigated.
However, all \cite{Zhu2014,Chen2015,Wang2015,Wanghm2016,Zhu12016,Wu2016} do not take channel estimation errors into consideration.
To the best of the authors' knowledge, the optimal solution of robust beamforming for the physical layer security of  massive MIMO systems has not been reported yet.

In this paper, we consider robust beamforming for the physical layer security of  multiuser massive MIMO systems following BMDA scheme \cite{BDMA1,BDMA2},
where the estimated channels are lied orthogonally to each other and the channel estimation errors are bounded by ball constants.
The proposed design utilizes simultaneous robust information and  artificial noise (AN) beamforming to provide the legal users and Eve with different signal-to-interference-and-noise ratio (SINR), meanwhile minimizing the transmit power of BS.
The resulted problem is a well-known NP-hard \cite{boyd2004} optimization, where the computational complexity \cite{Jiang2010}  increases exponentially with the number of antennas.
Consequently, existing robust beamforming algorithms for conventional communication systems \cite{Rong2006,Palomar2009,Vucic2009,Mostafa2016} cannot be applied  in massive MIMO systems due to their forbiddingly huge computational complexity.
Nevertheless, we demonstrate that under the BDMA massive MIMO scheme, the proposed design can be globally solved.
 More importantly, the optimal robust beamforming vectors can be derived in closed-form, which will greatly reduce the computational
complexity and is suitable for practical applications.
Interestingly, it is shown that the  AN beamforming is not necessary for the proposed robust optimization.

The rest of this paper is organized as follows:  Section~\ref{Sec.2}
describes the BDMA massive MIMO channel model and formulates the proposed robust design;
Section~\ref{Sec.3} converts the initial non-convex optimization into a convex semi-definite programming (SDP) and proves the optimality of semi-definite relaxation (SDR);
Closed-form solutions for the power allocation problem are derived in Section~\ref{Sec.4};
Simulation results are provided in Section~\ref{Sec.5}
and conclusions are drawn in Section~\ref{Sec.6}.

\emph{Notation:}  Vectors and matrices are boldface small and
capital letters, respectively;
The Hermitian, inverse and Moore-Penrose inverse of ${\bm A}$ are
denoted by ${\bm A}^{\rm{H}}$, ${\bm A}^{-1}$ and ${\bm A}^{\dag}$ respectively;
${\rm{Tr}}(\bm{A})$ defines the trace;
$\bm{I}$ and $\bm{0}$ represent an identity matrix and an all-zero matrix, respectively, with appropriate dimensions;
$\bm{A}\succeq \bm{0}$ and $\bm{A}\succ
\bm{0}$ mean that $\bm{A}$ is positive semi-definite and positive
definite, respectively;
The distribution of a circularly symmetric complex
Gaussian (CSCG) random variable with zero mean and variance
$\sigma^2$ is defined as $\mathcal{CN}(0,\sigma^2)$, and $\sim$
means ``distributed as"; $\mathbb{R}^{a\times b}$ and $\mathbb{C}^{a\times b}$ denote the spaces
of $a\times b$ matrices with real- and complex-valued entries, respectively;  $\|\bm{x}\|$ is
the Euclidean norm of a vector $\bm{x}$.

\section{System Model and Problem Formulation}\label{Sec.2}

\subsection{System Model}

\begin{figure}[t]
    \centering
    \includegraphics[scale=0.55]{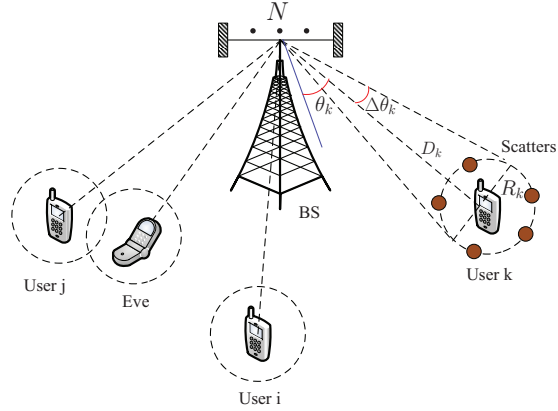}
    \caption{Multiuser massive MIMO channel model in the form of ULA.}\label{fig1}
\end{figure}

Let us consider physical layer security for a multiuser massive MIMO system shown in Fig.~\ref{fig1}. It is assumed that the BS is equipped with $N\gg 1$ antennas in the form of uniform linear array (ULA), where the antenna spacing is less than or equal to half wavelength.
There are $K$ legal users and one Eve randomly distributed in the coverage area, where all the legal users and the Eve are quipped with single antenna.
The channel from the $k$th user to BS can be expressed as \cite{Yin2013,Xie20163} $\bm{h}_k=\int_{\theta\in\Theta_k}\alpha_{k}(\theta)\bm{a}(\theta)d\theta$,
where $\Theta_k$ is the incoming angular spread of user-$k$ and $\alpha_{k}(\theta)$ is the corresponding spatial power spectrum.
Moreover, $\bm{a}(\theta)\in \mathbb{C}^{N\times 1}$ is the steering vector that can be expressed as $\bm{a}(\theta)=\left[1,e^{j\frac{2\pi d}{\lambda}\sin\theta},\ldots,e^{j\frac{2\pi d}{\lambda}(N-1)\sin\theta}\right]^{\rm H}$,
where $d$ is the antenna spacing and  $\lambda$ denotes the signal wavelength. Similarly, the channel from the Eve to BS can be expressed as $\bm{h}_e=\int_{\theta\in\Theta_e}\alpha_{e}(\theta)\bm{a}(\theta) d\theta$,
where $\Theta_e$ is the incoming angular spread of Eve and $\alpha_{e}(\theta)$ is the corresponding spatial power spectrum.

In this paper, we adopt the recently popular low complexity BDMA channel estimation scheme \cite{BDMA1,BDMA2} such that pilots of different users are transmitted through orthogonal spatial directions\footnote{In particular, the columns of discrete Fourier transform (DFT) matrix with non-overlap indices are assigned to different users as the training directions. Such a scheme is not   optimal in terms of training design but could be implemented with much higher efficiency. Moreover, as long as $N$ is large, which is the case of massive MIMO,  the performance loss of BDMA channel estimation is small compared to the optimal one.}, and the estimated channels for different users must be orthogonal, i.e.,
\begin{align}
\tilde{\bm{h}}_i^{\rm H}\tilde{\bm{h}}_j=0,\  \tilde{\bm{h}}_i^{\rm H}\tilde{\bm{h}}_e=0,\ \forall i\neq j,   \label{model.3}
\end{align}
Please refer to \cite{BDMA1,BDMA2} for detailed discussion of BDMA.
We  then assume that the real channel vectors $\bm{h}_k$ and $\bm{h}_e$ lie around the estimated channel vectors $\tilde{\bm{h}}_k$ and $\tilde{\bm{h}}_e$, respectively, i.e.,
\begin{align}
\bm{h}_k \in \mathcal{U}_k=\left\{\tilde{\bm{h}}_k+\bm{\delta}_k \ | \   \|\bm{\delta}_k\|\leq \epsilon_k \right\},\ \bm{h}_e \in \mathcal{U}_e=\left\{\tilde{\bm{h}}_e+\bm{\delta}_e \ | \   \|\bm{\delta}_e\|\leq \epsilon_e \right\}\label{2.1.2}
\end{align}
where $\bm{\delta}_k\in \mathbb{C}^{N\times 1}$ and $\bm{\delta}_e\in \mathbb{C}^{N\times 1}$ are the channel estimation errors whose norms are assumed to be bounded by $\epsilon_k$ and $\epsilon_e$ respectively.\footnote{Note that as long as $N$ is not infinity, the real channel $\bm{h}_k$ and $\bm{h}_e$ cannot be perfectly orthogonal.}

\subsection{Problem Formulation}

Let us  apply AN at BS during the downlink data transmission to provide the strongest distortion to Eve, i.e., the baseband signal from BS can be expressed as
\begin{align}
\bm{x}_b= \sum_{k=1}^{K}\bm{s}_k v_k+\bm{w}v_w,\label{2.2.3}
\end{align}
where $v_k\sim \mathcal{CN}(0,1)$ denotes the data symbol for the $k$th user; $v_w \sim \mathcal{CN}(0,1)$ is the AN signal for the Eve;
$\bm{s}_k \in \mathbb{C}^{N\times 1}$ and $\bm{w}\in \mathbb{C}^{N\times 1}$ are information beamforming vector and AN beamforming vector,
respectively.

The downlink signal at the $k$th user can be expressed as
\begin{align}
y_k= \bm{h}_k^{\rm H}\bm{s}_k v_k+\sum_{i\neq k,i\in \mathcal{K}}\bm{h}_k^{\rm H}\bm{s}_i v_i+\bm{h}_k^{\rm H}\bm{w}_e v_e+n_k,\label{2.2.4}
\end{align}
where $n_k\sim \mathcal{CN}(0,\sigma_{k}^2)$ represents the antenna noise of the $k$th user.
The downlink signal at the Eve can be expressed as
\begin{align}
y_e= \sum_{k=1}^{K}\bm{h}_e^{\rm H}\bm{s}_k v_k+\bm{h}_e^{\rm H}\bm{w}_e v_e+n_e,\label{2.2.5}
\end{align}
where $n_k\sim \mathcal{CN}(0,\sigma_{k}^2)$ represents the antenna noise of the Eve.
Then the secret rate for the $k$th user are expressed as
\begin{align}
R_{k}=&\log_2\left(1\!+\!\frac{\|\bm{h}_k^{\rm H}\bm{s}_k\|^2}{\displaystyle\sum_{i\neq k}\|\bm{h}_k^{\rm H}\bm{s}_i\|^2\!+\!\|\bm{h}_k^{\rm H}\bm{w}_e\|^2\!+\!\sigma_k^2}\right)  \!-\! \log_2\left(1\!+\!\frac{\|\bm{h}_e^{\rm H}\bm{s}_k\|^2}{\displaystyle\sum_{i\neq k}\|\bm{h}_e^{\rm H}\bm{s}_i\|^2\!+\!\|\bm{h}_e^{\rm H}\bm{w}_e\|^2\!+\!\sigma_e^2}\right) \nonumber\\
\geq & \log_2\left(1+\gamma_k\right)- \log_2\left(1+\gamma_{e,k}\right),\label{2.2.6}
\end{align}
where $\gamma_k$ is the receive SINR for the $k$th user, and $\gamma_{e,k}$ is the receive SINR for Eve when Eve aims to eavesdrop the $k$th user.  Note that when $\gamma_k>\gamma_{e,k}>0,\forall k\in \{1,\ldots,K\}$, we could always obtain non-zero secret rate for each users, which implies that the physical layer security   could be strictly guaranteed.

Our target is to design simultaneous information beamforming vectors $\{\bm{s}_k\}$ and AN beamforming vector $\bm{w}_e$ to
 provide legal users and the Eve with different SINRs, meanwhile minimizing the transmit power of BS.
Taking the channel estimation errors into consideration, the robust transmit beamforming design can be formulated as
\begin{align}
\mathbf{P1}:\quad\underset{\{\bm{s}_k\},\bm{w}_e}{\min}  &  \quad  \|\bm{w}_e\|^2+\sum_{k=1}^{K}\|\bm{s}_k\|^2 \label{p1.7}\\
\textrm{s.t.} & \quad  \frac{\|\bm{h}_e^{\rm H}\bm{s}_k\|^2}{\displaystyle\sum_{i\neq k}\|\bm{h}_e^{\rm H}\bm{s}_i\|^2+\|\bm{h}_e^{\rm H}\bm{w}_e\|^2+\sigma_e^2}\leq \gamma_{e,k},\  \bm{h}_e \in \mathcal{U}_e,\label{p1.8}\\
& \quad  \frac{\|\bm{h}_k^{\rm H}\bm{s}_k\|^2}{\displaystyle\sum_{i\neq k}\|\bm{h}_k^{\rm H}\bm{s}_i\|^2+\|\bm{h}_k^{\rm H}\bm{w}_e\|^2+\sigma_k^2}\geq \gamma_k,\ \forall \bm{h}_k \in \mathcal{U}_k,\label{p1.9}\\
& \quad k=1,2,\ldots,K,
\end{align}
where $\gamma_{e,k}>0$ is the maximum allowable SINR for Eve, and $\gamma_k>0$ is the desired SINR for the $k$th user.
Note that the optimal solutions of $\mathbf{P1}$ are very hard to obtain in conventional MIMO systems\footnote{The proposed design $\mathbf{P1}$ has been proven to be NP hard in conventional communication systems\cite{Vucic2009}, where it is very difficult to obtain the optimal beamforming vectors.}.
Nevertheless,  we will next show that $\mathbf{P1}$ could be globally solved utilizing property (\ref{model.3}) of BDMA  massive MIMO systems.

\section{Optimal Robust Beamforming}\label{Sec.3}

\subsection{Problem Reformulation}

The main difficulty of solving $\mathbf{P1}$ lies in the constraints (\ref{p1.8}) and (\ref{p1.9}).
Substituting (\ref{2.1.2}) into (\ref{p1.8}), we can equivalently rewrite (\ref{p1.8}) as
\begin{align}
\left\{ \begin{array}{ll}
\left(\tilde{\bm{h}}_e\!+\!\bm{\delta}_e\right)^{\rm H}\left(\displaystyle\sum_{i\neq k}\bm{s}_i\bm{s}_i^{\rm H}+\bm{w}_e\bm{w}_e^{\rm H}-\displaystyle\frac{1}{\gamma_{e,k}}\bm{s}_k\bm{s}_k^{\rm H}\right)\left(\tilde{\bm{h}}_e\!+\!\bm{\delta}_e\right)\!+\!\sigma_e^2\geq 0,\\
-\bm{\delta}_e^{\rm H}\bm{I}\bm{\delta}_e+\epsilon_e^2\geq 0,\quad k=1,2,\ldots,K.
\end{array} \right. \label{3.1.11}
\end{align}
Substituting (\ref{2.1.2}) into (\ref{p1.9}), we can equivalently rewrite (\ref{p1.9}) as
\begin{align}
\left\{ \begin{array}{ll}
\left(\tilde{\bm{h}}_k\!+\!\bm{\delta}_k\right)^{\rm H}\left(\displaystyle\frac{1}{\gamma_k}\bm{s}_k\bm{s}_k^{\rm H}\!-\!\displaystyle\sum_{i\neq k}\bm{s}_i\bm{s}_i^{\rm H}-\bm{w}_e\bm{w}_e^{\rm H}\right)\left(\tilde{\bm{h}}_k\!+\!\bm{\delta}_k\right)\!-\!\sigma_k^2\geq 0,\\
-\bm{\delta}_k^{\rm H}\bm{I}\bm{\delta}_k+\epsilon_k^2\geq 0,\quad k=1,2,\ldots,K.
\end{array} \right. \label{3.1.12}
\end{align}


According to the Lemma of S-Procedure \cite{Boyed1994}, we know that the constraints  in (\ref{3.1.11}) hold true if and only if there exists $\mu_{e,k}\geq 0,\ k=1,2,\ldots,K$ such that
\begin{align}
\left[ \begin{array}{ccc}
\bm{X}_{e,k}+\mu_{e,k}\bm{I} & \bm{X}_{e,k}\tilde{\bm{h}}_{e} \\
\tilde{\bm{h}}_e^{\rm H}\bm{X}_{e,k}^{\rm H} & \tilde{\bm{h}}_e^{\rm H}\bm{X}_{e,k}\tilde{\bm{h}}_e+\sigma_e^2-\mu_{e,k}\epsilon_e^2
\end{array} \right]\succeq \bm{0},
\end{align}
where for simplicity, we define $\bm{X}_{e,k}$ as
\begin{align}
\bm{X}_{e,k}=\displaystyle\sum_{i\neq k}\bm{s}_i\bm{s}_i^{\rm H}+\bm{w}_e\bm{w}_e^{\rm H}-\displaystyle\frac{1}{\gamma_{e,k}}\bm{s}_k\bm{s}_k^{\rm H}.
\end{align}
Meanwhile, the constraints in (\ref{3.1.12}) hold true if and only if there exists $\mu_{s,k}\geq 0,\ k=1,2,\ldots,K$ such that
\begin{align}
\left[ \begin{array}{ccc}
\bm{X}_{s,k}+\mu_{s,k}\bm{I} & \bm{X}_{s,k}\tilde{\bm{h}}_k \\
\tilde{\bm{h}}_k^{\rm H}\bm{X}_{s,k}^{\rm H} & \tilde{\bm{h}}_k^{\rm H}\bm{X}_{s,k}\tilde{\bm{h}}_k-\sigma_k^2-\mu_{s,k}\epsilon_k^2
\end{array} \right]\succeq \bm{0},
\end{align}
where for simplicity, we define $\bm{X}_{s,k}$ as
\begin{align}
\bm{X}_{s,k}=\displaystyle\frac{1}{\gamma_k}\bm{s}_k\bm{s}_k^{\rm H}\!-\!\displaystyle\sum_{i\neq k}\bm{s}_i\bm{s}_i^{\rm H}-\bm{w}_e\bm{w}_e^{\rm H}.
\end{align}

It is then clear that $\mathbf{P1}$ can be equivalently re-expressed as
\begin{align}
\mathbf{P1\!-\!EQV}:\
&\underset{\{\mu_{e,k}\},\{\mu_{s,k}\},\bm{w}_e,\{\bm{s}_k\}}{\min}    \quad  {\rm Tr}(\bm{W}_e)+\sum_{k=1}^{K}{\rm Tr}(\bm{S}_k) \label{p1.eqv.0}\\
\textrm{s.t.}   & \quad  \left[ \begin{array}{ccc}
\bm{X}_{e,k}+\mu_{e,k}\bm{I} & \bm{X}_{e,k}\tilde{\bm{h}}_e \\
\tilde{\bm{h}}_e^{\rm H}\bm{X}_{e,k}^{\rm H} & \tilde{\bm{h}}_e^{\rm H}\bm{X}_{e,k}\tilde{\bm{h}}_e+\sigma_e^2-\mu_{e,k}\epsilon_e^2
\end{array} \right]\succeq \bm{0},\label{p1.eqv.1}\\
& \quad  \left[ \begin{array}{ccc}
\bm{X}_{s,k}+\mu_{s,k}\bm{I} & \bm{X}_{s,k}\tilde{\bm{h}}_k \\
\tilde{\bm{h}}_k^{\rm H}\bm{X}_{s,k}^{\rm H} & \tilde{\bm{h}}_k^{\rm H}\bm{X}_{s,k}\tilde{\bm{h}}_k-\sigma_k^2-\mu_{s,k}\epsilon_k^2
\end{array} \right]\succeq \bm{0},\label{p1.eqv.2}\\
& \quad \bm{X}_{e,k}=\displaystyle\sum_{i\neq k}\bm{S}_i+\bm{W}_e-\displaystyle\frac{1}{\gamma_{e,k}}\bm{S}_k , \label{p1.eqv.3}\\
& \quad \bm{X}_{s,k}=\displaystyle\frac{1}{\gamma_k}\bm{S}_k-\displaystyle\sum_{i\neq k}\bm{S}_i-\bm{W}_e, \label{p1.eqv.4}\\
& \quad \mu_{e,k}\geq 0,\quad \mu_{s,k}\geq 0,  \label{p1.eqv.5}\\
& \quad  \bm{W}_e=\bm{w}_e\bm{w}_e^{\rm H},\quad \bm{S}_k=\bm{s}_k\bm{s}_k^{\rm H},\quad k=1,2,\ldots,K,   \label{p1.eqv.6}
\end{align}
where $\{\mu_{e,k}\}$ and $\{\mu_{s,k}\}$ are the auxiliary variables generated by the S-Procedure.
Note that the nonlinear constraints in (\ref{p1.eqv.6}) are equivalent to:
\begin{align}
\bm{W}_e\succeq \bm{0},\quad \bm{S}_k\succeq \bm{0},\quad {\rm {Rank}}(\bm{W}_e)=1,\quad {\rm {Rank}}(\bm{S}_k)=1.
\end{align}
However, $\mathbf{P1\!-\!EQV}$ is still very hard to solve since the rank constraint in (\ref{p1.eqv.6}) is non-convex.
Dropping the the rank constraints ${\rm {Rank}}(\bm{W}_e)=1$ and ${\rm {Rank}}(\bm{S}_k)=1$ (the SDR technique \cite{luo2010}), we can obtain the following relaxed convex optimization:
\begin{align}
\mathbf{P1\!-\!SDR}:\
&\underset{\{\mu_{e,k}\},\{\mu_{s,k}\},\bm{W}_e,\{\bm{S}_k\}}{\min}    \quad  {\rm Tr}(\bm{W}_e)+\sum_{k=1}^{K}{\rm Tr}(\bm{S}_k) \label{p1.adr.1}\\
\textrm{s.t.}   & \quad  (\ref{p1.eqv.1})\sim (\ref{p1.eqv.5}),\\
& \quad \bm{W}_e\succeq \bm{0},\quad \bm{S}_k\succeq \bm{0},\quad k=1,2,\ldots,K,   \label{p1.sdr.6}
\end{align}
which is a semi-definite programming (SDP) problem that can be efficiently solved by the standard convex optimization tools \cite{boyd2004}.
However, the SDR technique does not guarantee rank-one solutions for the relaxed optimization.
Nevertheless, we will next show that $\mathbf{P1\!-\!SDR}$ indeed guarantees rank-one solutions for BDMA massive MIMO systems, i.e., $\mathbf{P1\!-\!SDR}$ is equivalent to $\mathbf{P1\!-\!EQV}$.

\subsection{Optimal Rank-one Solutions}

The following propositions  will be useful throughout the rest of the paper.

\emph{Proposition 1:} Suppose $\{\mu_{e,k}^\star\}$ and $\{\mu_{s,k}^\star\}$ are the optimal auxiliary variables of $\mathbf{P1\!-\!SDR}$. There must be

(a) $\mu_{e,k}^\star>0,\ \forall k\in \{1,\ldots,K\}$;
(b) $\mu_{s,k}^\star>0,\ \forall k\in \{1,\ldots,K\}$.
\begin{proof}
See Appendix~\ref{Appen:A}.
\end{proof}

\emph{Proposition 2:} Suppose $\bm{W}_e^\star$ and $\{\bm{S}_k^\star\}$ are the optimal transmit covariances of $\mathbf{P1\!-\!SDR}$. There must be

(a) ${\rm Rank}\left(\bm{X}_{e,k}^\star+\mu_{e,k}^\star\bm{I}\right)\geq 1$, $\forall k\in \{1,\ldots,K\}$;
(b) ${\rm Rank}\left(\bm{X}_{s,k}^\star+\mu_{s,k}^\star\bm{I}\right)\geq 1$, $\forall k\in \{1,\ldots,K\}$.
\begin{proof}
See Appendix~\ref{Appen:B}.
\end{proof}

\emph{Remark 1:} Proposition 1 and Proposition 2  indicate that the optimal AN and information transmit covariance could not be full-rank, i.e., ${\rm Rank}\left(\bm{W}_{e}^\star\right)< N$ and ${\rm Rank}\left(\bm{S}_{k}^\star\right)< N,\forall k\in\{1,\ldots,K\}$.
Moreover, from (\ref{Appen.b.8}) in Appendix~\ref{Appen:B}, we know that ${\rm Rank}\left(\bm{W}_{e}^\star\right)$ and ${\rm Rank}\left(\bm{S}_{k}^\star\right)$ can be further constrained as ${\rm Rank}\left(\bm{W}_{e}^\star\right)\leq K+1$ and ${\rm Rank}\left(\bm{S}_{k}^\star\right)\leq K+1,\forall k\in\{1,\ldots,K\}$.

From Proposition 1 and Proposition 2, we can reformulate $\mathbf{P1\!-\!SDR}$ to gain more insightful solutions.
Let the eigenvalue decomposition (EVD) of $\bm{X}_{e,k}$ be $\bm{X}_{e,k}=\bm{U}_{e,k}\bm{\Lambda}_{e,k}\bm{U}_{e,k}^{\rm H}$ with the eigenvalues $q_{e,k,1}\geq\ldots\geq q_{e,k,N}$,
and the EVD of $\bm{X}_{s,k}$ be $\bm{X}_{s,k}=\bm{U}_{s,k}\bm{\Lambda}_{s,k}\bm{U}_{s,k}^{\rm H}$ with the eigenvalues $q_{s,k,1}\geq\ldots\geq q_{s,k,N}$.
Due to the fact that $\mu_{e,k}\bm{I}=\bm{U}_{e,k}\left(\mu_{e,k}\bm{I}\right)\bm{U}_{e,k}^{\rm H}$ and $\mu_{s,k}\bm{I}=\bm{U}_{s,k}\left(\mu_{s,k}\bm{I}\right)\bm{U}_{s,k}^{\rm H}$, the EVDs of $\bm{X}_{e,k}+\mu_{e,k}\bm{I}$ and $\bm{X}_{s,k}+\mu_{s,k}\bm{I}$ can be represented as
\begin{align}
\bm{X}_{e,k}+\mu_{e,k}\bm{I}=\bm{U}_{e,k}\left(\bm{\Lambda}_{e,k}+\mu_{e,k}\bm{I}\right)\bm{U}_{e,k}^{\rm H},\quad
\bm{X}_{s,k}+\mu_{s,k}\bm{I}=\bm{U}_{s,k}\left(\bm{\Lambda}_{s,k}+\mu_{s,k}\bm{I}\right)\bm{U}_{s,k}^{\rm H},
\end{align}
respectively, where $\bm{\Lambda}_{e,k}+\mu_{e,k}\bm{I}$ is diagonal with the eigenvalues $q_{e,k,1}+\mu_{e,k}\geq\ldots\geq q_{e,k,N}+\mu_{e,k}$ and $\bm{\Lambda}_{s,k}+\mu_{s,k}\bm{I}$ is diagonal with the eigenvalues $q_{s,k,1}+\mu_{s,k}\geq\ldots\geq q_{s,k,N}+\mu_{s,k}$.
Using Proposition~1 and Proposition~2, it is easily known that $q_{e,k,1}+\mu_{e,k}>0$ and $q_{s,k,1}+\mu_{s,k}>0$ hold.
Then assuming ${\rm Rank}\left(\bm{X}_{e,k}+\mu_{e,k}\bm{I}\right)=l_{e,k}\geq 1$ and ${\rm Rank}\left(\bm{X}_{s,k}+\mu_{s,k}\bm{I}\right)=l_{s,k}\geq 1$, there are
\begin{align}
q_{e,k,i}+\mu_{e,k}=0, \forall i\in \{l_{e,k}+1,\ldots,N\}, \forall k\in \{1,\ldots,K\}, \label{Schur.17}\\
q_{s,k,i}+\mu_{s,k}=0, \forall i\in \{l_{s,k}+1,\ldots,N\}, \forall k\in \{1,\ldots,K\}.
\end{align}

Define $\bm{\Sigma}_{e,k,+}$ and $\bm{\Sigma}_{s,k,+}$  as
\begin{align}
\bm{\Sigma}_{e,k,+}=\left[ \begin{array}{ccc}
q_{e,k,1}+\mu_{e,k} & \bm{0} & 0\\
\bm{0} & \ddots & \bm{0}\\
0     &  \bm{0}      & q_{e,k,l_{e,k}}+\mu_{e,k}
\end{array} \right],
\bm{\Sigma}_{s,k,+}=\left[ \begin{array}{ccc}
q_{s,k,1}+\mu_{s,k} & \bm{0} & 0\\
\bm{0} & \ddots & \bm{0}\\
0     &  \bm{0}  & q_{s,k,l_{e,k}}+\mu_{s,k}
\end{array} \right],\nonumber
\end{align}
respectively.
The Moore-Penrose inverses \cite{Penrose1955} of $\bm{X}_{e,k}+\mu_{e,k}\bm{I}$ and $\bm{X}_{s,k}+\mu_{s,k}\bm{I}$ can be derived as\footnote{Note that if ${\rm Rank}\left(\bm{X}_{e,k}+\mu_{e,k}\bm{I}\right)=N$ and ${\rm Rank}\left(\bm{X}_{s,k}+\mu_{s,k}\bm{I}\right)=N$, then there are $\left(\bm{X}_{e,k}+\mu_{e,k}\bm{I}\right)^{\dag}=\left(\bm{X}_{e,k}+\mu_{e,k}\bm{I}\right)^{-1}$ and $\left(\bm{X}_{s,k}+\mu_{s,k}\bm{I}\right)^{\dag}=\left(\bm{X}_{s,k}+\mu_{s,k}\bm{I}\right)^{-1}$.}
\begin{align}
\left(\bm{X}_{e,k}+\mu_{e,k}\bm{I}\right)^{\dag}=\bm{U}_{e,k}\left[ \begin{array}{ccc}
\bm{\Sigma}^{-1}_{e,k,+} & \bm{0} \\
\bm{0} & \bm{0}
\end{array} \right]\bm{U}_{e,k}^{\rm H},
\left(\bm{X}_{s,k}+\mu_{s,k}\bm{I}\right)^{\dag}=\bm{U}_{s,k}\left[ \begin{array}{ccc}
\bm{\Sigma}^{-1}_{s,k,+} & \bm{0} \\
\bm{0} & \bm{0}
\end{array} \right]\bm{U}_{s,k}^{\rm H}, \label{3.2.33}
\end{align}
respectively. Then we provide the following lemma.

\emph{Lemma 2 (Generalized Schur's Complement \cite{Horn1985}):} Let $\bm{M}=\left[\bm{A}, \bm{B}; \bm{B}^{\rm H}, \bm{C}  \right]$
be a Hermitian matrix. Then, $\bm{M}\succeq \bm{0}$ if and only if $\bm{C}-\bm{B}^{\rm H}\bm{A}^{\dag}\bm{B}\succeq \bm{0}$ and $\left(\bm{I}-\bm{A}\bm{A}^{\dag}\right)\bm{B}=\bm{0}$ (assuming $\bm{A}\succeq \bm{0}$),
or $\bm{A}-\bm{B}\bm{C}^{\dag}\bm{B}^{\rm H}\succeq \bm{0}$ and $\left(\bm{I}-\bm{C}\bm{C}^{\dag}\right)\bm{B}^{\rm H}=\bm{0}$ (assuming $\bm{C}\succeq \bm{0}$), where $\bm{A}^{\dag}$ and $\bm{C}^{\dag}$ are the  generalized inverses of $\bm{A}$ and $\bm{C}$, respectively.

Using Lemma 2, we know that $\mathbf{P1\!-\!SDR}$ could be equivalently rewritten as
\begin{align}
 \mathbf{P1\!-\!SDR\!-\!EQV}:&\nonumber\\
\underset{\{\mu_{e,k}\},\{\mu_{s,k}\},\bm{W}_e,\{\bm{S}_k\}}{\min}  &  \quad  {\rm Tr}(\bm{W}_e)+\sum_{k=1}^{K}{\rm Tr}(\bm{S}_k) \label{p1.sdr.eqv.1}\\
\textrm{s.t.}  &  \quad \bm{X}_{e,k}+\mu_{e,k}\bm{I}\succeq \bm{0},\quad \bm{X}_{s,k}+\mu_{s,k}\bm{I}\succeq \bm{0},\label{p1.sdr.eqv.2a}\\
&  \quad \left[\bm{I}-\left(\bm{X}_{e,k}+\mu_{e,k}\bm{I}\right)\left(\bm{X}_{e,k}+\mu_{e,k}\bm{I}\right)^{\dag}\right]\bm{X}_{e,k}\tilde{\bm{h}}_e=\bm{0},\label{p1.sdr.eqv.2}\\
& \quad \tilde{\bm{h}}_e^{\rm H}\bm{X}_{e,k}\tilde{\bm{h}}_e+\sigma_e^2-\mu_{e,k}\epsilon_e^2 -\tilde{\bm{h}}_e^{\rm H}\bm{X}_{e,k}^{\rm H}\left(\bm{X}_{e,k}+\mu_{e,k}\bm{I}\right)^{\dag}\bm{X}_{e,k}\tilde{\bm{h}}_e\geq 0,\label{p1.sdr.eqv.3}\\
& \quad  \left[\bm{I}-\left(\bm{X}_{s,k}+\mu_{s,k}\bm{I}\right)\left(\bm{X}_{s,k}+\mu_{s,k}\bm{I}\right)^{\dag}\right]\bm{X}_{s,k}\tilde{\bm{h}}_k=\bm{0},\label{p1.sdr.eqv.4}\\
& \quad \tilde{\bm{h}}_k^{\rm H}\bm{X}_{s,k}\tilde{\bm{h}}_k-\sigma_k^2-\mu_{s,k}\epsilon_k^2 -\tilde{\bm{h}}_k^{\rm H}\bm{X}_{s,k}^{\rm H}\left(\bm{X}_{s,k}+\mu_{s,k}\bm{I}\right)^{\dag}\bm{X}_{s,k}\tilde{\bm{h}}_k\geq 0,\label{p1.sdr.eqv.5}\\
& \quad \bm{X}_{e,k}=\displaystyle\sum_{i\neq k}\bm{S}_i+\bm{W}_e-\displaystyle\frac{1}{\gamma_{e,k}}\bm{S}_k, \label{p1.sdr.eqv.6}\\
& \quad \bm{X}_{s,k}=\displaystyle\frac{1}{\gamma_k}\bm{S}_k-\displaystyle\sum_{i\neq k}\bm{S}_i-\bm{W}_e, \label{p1.sdr.eqv.7}\\
& \quad \mu_{e,k}\geq 0,\quad \mu_{s,k}\geq 0,\quad \bm{W}_e\succeq \bm{0},\quad \bm{S}_k\succeq \bm{0},\quad k=1,2,\ldots,K,   \label{p1.sdr.eqv.8}
\end{align}
where the constraints in (\ref{p1.sdr.eqv.2a})$\sim$(\ref{p1.sdr.eqv.5}) are derived from (\ref{p1.eqv.1}) and (\ref{p1.eqv.2}).

Next, we will investigate the constraints  (\ref{p1.sdr.eqv.2a})$\sim$(\ref{p1.sdr.eqv.5}) to obtain more insightful solutions.
Due to the fact that
\begin{align}
\left(\bm{X}_{e,k}+\mu_{e,k}\bm{I}\right)\left(\bm{X}_{e,k}+\mu_{e,k}\bm{I}\right)^{\dag}=\bm{U}_{e,k}
\left[ \begin{array}{ccc}
\bm{I}^{l_{e,k}\times l_{e,k}} & \bm{0} \\
\bm{0} & \bm{0}^{(N-l_{e,k})\times (N-l_{e,k})}
\end{array} \right]\bm{U}_{e,k}^{\rm H},\nonumber
\end{align}
Eq. (\ref{p1.sdr.eqv.2}) can be expressed as
\begin{align}
&\left[\bm{I}-\left(\bm{X}_{e,k}+\mu_{e,k}\bm{I}\right)\left(\bm{X}_{e,k}+\mu_{e,k}\bm{I}\right)^{\dag}\right]\bm{X}_{e,k}\tilde{\bm{h}}_e\nonumber\\
=& \bm{U}_{e,k}\left[ \begin{array}{ccc}
\bm{0}^{l_{e,k}\times l_{e,k}} & \bm{0} \\
\bm{0} & \bm{I}^{(N-l_{e,k})\times (N-l_{e,k})}
\end{array} \right]\bm{U}_{e,k}^{\rm H}\bm{U}_{e,k}\left[ \begin{array}{ccc}
\bm{\Lambda}_{e,k}^{l_{e,k}\times l_{e,k}} & \bm{0} \\
\bm{0} & \bm{\Lambda}_{e,k}^{(N-l_{e,k})\times (N-l_{e,k})}
\end{array} \right]\bm{U}_{e,k}^{\rm H}\tilde{\bm{h}}_e\nonumber\\
=& \bm{U}_{e,k}\left[ \begin{array}{ccc}
\bm{0} & \bm{0} \\
\bm{0} & \bm{\Lambda}_{e,k}^{(N-l_{e,k})\times (N-l_{e,k})}
\end{array} \right]\bm{U}_{e,k}^{\rm H}\tilde{\bm{h}}_e=\bm{0},\label{3.2.42}
\end{align}
which says that $\bm{\Lambda}_{e,k}^{(N-l_{e,k})\times (N-l_{e,k})}\left[\bm{u}_{e,k,l_{e,k}+1}^{\rm H},\ldots,\bm{u}_{e,k,N}^{\rm H}\right]\tilde{\bm{h}}_e=\bm{0}$ holds true, where  $\bm{u}_{e,k,i}$ is the $i$th column of $\bm{U}_{e,k}$.
Similarly, Eq. (\ref{p1.sdr.eqv.4}) can be expressed as
\begin{align}
&\left[\bm{I}-\left(\bm{X}_{s,k}+\mu_{s,k}\bm{I}\right)\left(\bm{X}_{s,k}+\mu_{s,k}\bm{I}\right)^{\dag}\right]\bm{X}_{s,k}\tilde{\bm{h}}_k\nonumber\\
=& \bm{U}_{s,k}\left[ \begin{array}{ccc}
\bm{0}^{l_{s,k}\times l_{s,k}} & \bm{0} \\
\bm{0} & \bm{I}^{(N-l_{s,k})\times (N-l_{s,k})}
\end{array} \right]\bm{U}_{s,k}^{\rm H}\bm{U}_{s,k}\left[ \begin{array}{ccc}
\bm{\Lambda}_{s,k}^{l_{s,k}\times l_{s,k}} & \bm{0} \\
\bm{0} & \bm{\Lambda}_{s,k}^{(N-l_{s,k})\times (N-l_{s,k})}
\end{array} \right]\bm{U}_{s,k}^{\rm H}\tilde{\bm{h}}_k\nonumber\\
=& \bm{U}_{s,k}\left[ \begin{array}{ccc}
\bm{0} & \bm{0} \\
\bm{0} & \bm{\Lambda}_{s,k}^{(N-l_{s,k})\times (N-l_{s,k})}
\end{array} \right]\bm{U}_{s,k}^{\rm H}\tilde{\bm{h}}_k=\bm{0},\label{3.2.43}
\end{align}
which implies that $\bm{\Lambda}_{s,k}^{(N-l_{s,k})\times (N-l_{s,k})}\left[\bm{u}_{s,k,l_{s,k}+1}^{\rm H},\ldots,\bm{u}_{s,k,N}^{\rm H}\right]\tilde{\bm{h}}_k=\bm{0}$ holds true,
where $\bm{u}_{s,k,i}$ is the $i$th column of $\bm{U}_{s,k}$.

Due to $\bm{X}_{e,k}=\bm{X}_{e,k}^{\rm H}$, Eq. (\ref{p1.sdr.eqv.3}) can be expressed as
\begin{align}
& \tilde{\bm{h}}_e^{\rm H}\bm{X}_{e,k}\tilde{\bm{h}}_e+\sigma_e^2-\mu_{e,k}\epsilon_e^2 -\tilde{\bm{h}}_e^{\rm H}\bm{X}_{e,k}^{\rm H}\left(\bm{X}_{e,k}+\mu_{e,k}\bm{I}\right)^{\dag}\bm{X}_{e,k}\tilde{\bm{h}}_e   \nonumber\\
= & {\rm Tr}\left(\tilde{\bm{H}}_e\bm{X}_{e,k}\right)+\sigma_e^2-\mu_{e,k}\epsilon_e^2 -{\rm Tr}\left[\tilde{\bm{H}}_e\bm{X}_{e,k}^{\rm H}\left(\bm{X}_{e,k}+\mu_{e,k}\bm{I}\right)^{\dag}\bm{X}_{e,k}\right] \nonumber\\
= & {\rm Tr}\left\{\left(\tilde{\bm{H}}_e\bm{X}_{e,k}\right)\left[\bm{I} -\left(\bm{X}_{e,k}+\mu_{e,k}\bm{I}\right)^{\dag}\bm{X}_{e,k}\right]\right\}+\sigma_e^2-\mu_{e,k}\epsilon_e^2 \geq 0,\label{3.2.44}
\end{align}
where $\tilde{\bm{H}}_e=\tilde{\bm{h}}_e\tilde{\bm{h}}_e^{\rm H}$. Similarly, Eq. (\ref{p1.sdr.eqv.5}) can be expressed as
\begin{align}
& \tilde{\bm{h}}_k^{\rm H}\bm{X}_{s,k}\tilde{\bm{h}}_k-\sigma_k^2-\mu_{s,k}\epsilon_k^2 -\tilde{\bm{h}}_k^{\rm H}\bm{X}_{s,k}^{\rm H}\left(\bm{X}_{s,k}+\mu_{s,k}\bm{I}\right)^{\dag}\bm{X}_{s,k}\tilde{\bm{h}}_k   \nonumber\\
= & {\rm Tr}\left(\tilde{\bm{H}}_k\bm{X}_{s,k}\right)-\sigma_k^2-\mu_{s,k}\epsilon_k^2 -{\rm Tr}\left[\tilde{\bm{H}}_k\bm{X}_{s,k}^{\rm H}\left(\bm{X}_{s,k}+\mu_{s,k}\bm{I}\right)^{\dag}\bm{X}_{s,k}\right] \nonumber\\
= & {\rm Tr}\left\{\left(\tilde{\bm{H}}_k\bm{X}_{s,k}\right)\left[\bm{I} -\left(\bm{X}_{s,k}+\mu_{s,k}\bm{I}\right)^{\dag}\bm{X}_{s,k}\right]\right\}-\sigma_k^2-\mu_{s,k}\epsilon_k^2 \geq 0,\label{3.2.45}
\end{align}
where $\tilde{\bm{H}}_k=\tilde{\bm{h}}_k\tilde{\bm{h}}_k^{\rm H}$.

\emph{Theorem 1:} When Eq. (\ref{model.3}) holds, the optimal solutions of $\mathbf{P1\!-\!SDR\!-\!EQV}$ must satisfy
\begin{enumerate}
\item [(a)] ${\rm Rank}\left(\bm{W}_e^\star\right)\leq 1$,
where $\bm{W}_e^\star=\bm{w}_e^{\star}\bm{w}_e^{\star{\rm H}}=P^{\star}_{e}\tilde{\bm{h}}_{e}\tilde{\bm{h}}_e^{\rm H}/\|\tilde{\bm{h}}_{e}\|^2$ is the optimal AN transmit covariance that can be derived through the closed-form AN beamforming vector, and $P^{\star}_{e}$ is the power allocated for the AN beamforming;
\item [(b)] ${\rm Rank}\left(\bm{S}_k^\star\right)\leq 1, \ \forall k\in \{1,\ldots,K\}$,
where $\bm{S}_k^\star=\bm{s}_k^{\star}\bm{s}_k^{\star{\rm H}}=P^{\star}_{k}\tilde{\bm{h}}_{k}\tilde{\bm{h}}_k^{\rm H}/\|\tilde{\bm{h}}_{k}\|^2$ is the optimal information transmit covariance that can be derived through the closed-form information beamforming vectors, and $P^{\star}_{k}$ is the power allocated for the $k$th information beamforming.
\end{enumerate}

\begin{proof}
See Appendix~\ref{Appen:C}.
\end{proof}

\emph{Remark 2:} Theorem 1 indicates that simultaneous AN- and information-beamforming is the optimal transmit strategy for $\mathbf{P1\!-\!SDR\!-\!EQV}$ rather than the joint precoding scheme. Thus, the SDR solutions are indeed optimal for $\mathbf{P1}$.

\section{Optimal Power Allocation}\label{Sec.4}

In Section~\ref{Sec.3}, it is shown that the optimal solutions of $\mathbf{P1}$ could be derived by solving $\mathbf{P1\!-\!SDR\!-\!EQV}$, where the optimal beamforming vectors are derived in closed-form. However, the optimal power allocations $P_e^\star$ and $\{P_k^\star\}$ remain  unknown.
Since the optimal AN-beamforming and information-beamforming vectors are $\bm{w}_e^\star=\sqrt{P^{\star}_{e}}\tilde{\bm{h}}_{e}/\|\tilde{\bm{h}}_{e}\|$ and $\bm{s}_k^\star=\sqrt{P^{\star}_{k}}\tilde{\bm{h}}_{k}/\|\tilde{\bm{h}}_{k}\|$,
$\mathbf{P1}$ can be simplified to
\begin{align}
\mathbf{P2}:\
\underset{\{\mu_{e,k}\},\{\mu_{s,k}\},P_e,\{P_{k}\}}{\min}  &  \quad  P_e+\sum_{k=1}^{K}P_{k} \label{p2.1}\\
\textrm{s.t.}\kern10pt   & \quad \frac{\mu_{e,k} P_{e} \|\tilde{\bm{h}}_{e}\|^2}{\mu_{e,k}+P_{e}}+\sigma_e^2-\mu_{e,k}\epsilon_e^2\geq 0,\label{p2.2}\\
& \quad \frac{\mu_{s,k} P_{{k}} \|\tilde{\bm{h}}_{k}\|^2}{\mu_{s,k}\gamma_k+P_{{k}}}-\sigma_k^2-\mu_{s,k}\epsilon_k^2\geq 0,\label{p2.3}\\
& \quad \mu_{e,k}>0,\ \mu_{s,k}> 0,\ P_e\geq 0,\ P_{k}\geq 0, \  k=1,2,\ldots,K. \label{p2.4}
\end{align}
Define
\begin{align}
f_{e,k}(\mu_{e,k},P_{e})\triangleq\frac{\mu_{e,k} P_{e} \|\tilde{\bm{h}}_{e}\|^2}{\mu_{e,k}+P_{e}},\quad f_{s,k}(\mu_{s,k},P_{k})\triangleq\frac{\mu_{s,k} P_{k} \|\tilde{\bm{h}}_{k}\|^2}{\mu_{s,k}\gamma_k+p_{k}},
\end{align}
whose Hessian matrixes are given by
\begin{align}
&\nabla^2f_{e,k}(\mu_{e,k},P_{e})=\frac{-2\|\tilde{\bm{h}}_{e}\|^2}{(\mu_{e,k}+P_{e})^3}\left[ \begin{array}{ccc}
P_e^{2} & -\mu_{e,k} P_{e} \\
-\mu_{e,k}P_{e} & \mu_{e,k}^2
\end{array} \right]\preceq \bm{0},\\
&\nabla^2f_{s,k}(\mu_{s,k},P_{k})=\frac{-2\gamma_k\|\tilde{\bm{h}}_{k}\|^2}{(\mu_{s,k}\gamma_k+P_{k})^3}\left[ \begin{array}{ccc}
P_{k}^{2} & -\mu_{s,k} P_{k} \\
-\mu_{s,k}P_{k} & \mu_{s,k}^2
\end{array} \right]\preceq \bm{0},
\end{align}
respectively, which implies  that $f_{e,k}(\mu_{e,k},P_{e})$ and $f_{s,k}(\mu_{s,k},P_{k})$ are concave functions. Thus, $\mathbf{P2}$ is a convex optimization.

\subsection{Optimal AN Power Allocation}

The optimal AN power allocation $P^{\star}_{e}$ can be summarized in the following theorem.

\emph{Theorem 2:} At the optimal point, the AN power allocation $P^{\star}_{e}$ of $\mathbf{P2}$ must satisfy
\begin{align}
P^{\star}_{e}=0,
\end{align}
where the corresponding auxiliary variables $\{\mu_{e,k}\}$ satisfy $\mu_{e,k}=\sigma_e^2/\epsilon_e^2>0,\forall k\in \{1,\ldots,K\}$.

\begin{proof}
Let us show that $P^{\star}_{e}=0$ from contradiction. Assume $\{\mu_{e,k}^{\star}\}$, $\{\mu_{s,k}^{\star}\}$, $P_e^{\star}$ and $\{P_{k}^{\star}\}$ are the optimal solutions of $\mathbf{P2}$, where $P_e^{\star}>0$ is satisfied. Then we can provide a group of new solutions:
\begin{align}
\mu_{e,k}^*=\frac{\sigma_e^2}{\epsilon_e^2},\quad \mu_{s,k}^{*}=\mu_{s,k}^{\star},\quad P^{*}_{e}=0,\quad P_{k}^{*}=P_{k}^{\star},\quad  \forall k\in \{1,\ldots,K\}.
\end{align}
Substituting $\{\mu_{e,k}^{*}\}$, $\{\mu_{s,k}^{*}\}$, $P_e^{*}$ and $\{P_{k}^{*}\}$ into $\mathbf{P2}$, we obtain
\begin{align}
&  P_e^*+\sum_{k=1}^{K}P_{k}^*=0+\sum_{k=1}^{K}P_{k}^\star< P_e^\star+\sum_{k=1}^{K}P_{k}^\star    \label{PP2.1}\\
& \frac{\mu_{e,k}^* P_{e}^* \|\tilde{\bm{h}}_{e}\|^2}{\mu_{e,k}^*+P_{e}^*}+\sigma_e^2-\mu_{e,k}^*\epsilon_e^2= 0,\label{PP2.2}\\
& \frac{\mu_{s,k}^* P_{{k}}^* \|\tilde{\bm{h}}_{k}\|^2}{\mu_{s,k}^*\gamma_k+P_{k}^*}-\sigma_k^2-\mu_{s,k}^*\epsilon_k^2\geq 0,\label{PP2.3}\\
& \mu_{e,k}^*=\frac{\sigma_e^2}{\epsilon_e^2}>0,\ \mu_{s,k}^{*}=\mu_{s,k}^{\star}> 0,\ P_e^*= 0,\ P_{k}^*=P_{k}^{\star}\geq 0, \  k=1,2,\ldots,K. \label{PP2.4}
\end{align}
From  (\ref{PP2.2})$\sim$(\ref{PP2.4}), we conclude that  $\{\mu_{e,k}^{*}\}$, $\{\mu_{s,k}^{*}\}$, $P_e^{*}$ and $\{P_{k}^{*}\}$ satisfy all the constraints of $\mathbf{P2}$.
While we know from (\ref{PP2.2}) that $\{\mu_{e,k}^{*}\}$, $\{\mu_{s,k}^{*}\}$, $P_e^{*}$ and $\{P_{k}^{*}\}$ will always provide smaller object value, which contradicts the assumption that  $\{\mu_{e,k}^{\star}\}$, $\{\mu_{s,k}^{\star}\}$, $P_e^{\star}$ and $\{P_{k}^{\star}\}$ are the optimal solutions. Thus, at the optimal point there must be $P^{\star}_{e}=0$.
\end{proof}

Theorem 2 says that in order to optimally guarantee the the physical layer security for BDMA massive MIMO system,
one should not adopt the AN beamforming, which is much different from the conventional MIMO case.
\subsection{Optimal Information Power Allocation}
The Lagrange of $\mathbf{P2}$ is defined as
\begin{align}
  &\mathcal{L}\left(\{\mu_{e,k}\},\{\mu_{s,k}\},P_e,\{P_{k}\},\{\varpi_k\},\{\xi_k\}\right)=P_e+\sum_{k=1}^{K}P_{k}\nonumber\\
  & -\sum_{k=1}^{K}\left[\varpi_k\left(\frac{\mu_{e,k} P_{e} \|\tilde{\bm{h}}_{e}\|^2}{\mu_{e,k}+P_{e}}+\sigma_e^2-\mu_{e,k}\epsilon_e^2\right) \right] -\sum_{k=1}^{K}\left[\xi_k\left(\frac{\mu_{s,k} P_{k} \|\tilde{\bm{h}}_{k}\|^2}{\mu_{s,k}\gamma_k+P_{k}}-\sigma_k^2-\mu_{s,k}\epsilon_k^2\right)\right],
\end{align}
where $\{\varpi_k\geq 0\}$ and $\{\xi_k\geq 0\}$ are the dual variables associated with the constraints  (\ref{p2.2}) and (\ref{p2.3}), respectively.
The Karush-Kuhn-Tucker (KKT) conditions related to $\{\mu_{e,k}\}$ $\{\mu_{s,k}\}$, $P_e$ and $\{P_{k}\}$ can be formulated as $ \forall  k\in \{1,\ldots,K\}$
\begin{align}
&\frac{\partial \mathcal{L}}{\partial \mu_{e,k}^\star}=\varpi^\star_k\epsilon_e^2-\frac{\varpi_k^\star P_{e}^{\star 2}\|\tilde{\bm{h}}_{e}\|^2}{(\mu_{e,k}^\star+P_{e}^\star)^2}=0, \label{kkt.1}\\
&\frac{\partial \mathcal{L}}{\partial \mu_{s,k}^\star}=\xi_k^\star\epsilon_k^2-\frac{\xi_k^\star P_{k}^{\star 2}\|\tilde{\bm{h}}_{k}\|^2}{(\mu_{s,k}^\star\gamma_k+P_{k}^\star)^2}=0, \label{kkt.2}\\
&\frac{\partial \mathcal{L}}{\partial P_{e}^\star}=1-\sum_{k=1}^{K}\frac{\varpi_k^\star\mu_{e,k}^{\star 2}\|\tilde{\bm{h}}_{e}\|^2}{(\mu_{e,k}^\star+P_{e}^\star)^2}=0, \label{kkt.3}\\
&\frac{\partial \mathcal{L}}{\partial P_{k}^\star}=1-\frac{\xi_k^\star\mu_{s,k}^{\star 2}\gamma_k\|\tilde{\bm{h}}_{k}\|^2}{(\mu_{s,k}^\star\gamma_k+P_{k}^\star)^2}=0, \label{kkt.4}\\
&\varpi_k^\star\left(\frac{\mu_{e,k}^\star P_{e}^\star \|\tilde{\bm{h}}_{e}\|^2}{\mu_{e,k}^\star+P_{e}^\star}+\sigma_e^2-\mu_{e,k}^\star\epsilon_e^2\right)=0, \label{kkt.5} \\
&\xi_k^\star\left(\frac{\mu_{s,k}^\star P_{k}^\star \|\tilde{\bm{h}}_{k}\|^2}{\mu_{s,k}^\star\gamma_k+P_{k}^\star}-\sigma_k^2-\mu_{s,k}^\star\epsilon_k^2\right)=0, \label{kkt.6}
\end{align}
where $\{\mu_{e,k}^\star>0\}$, $\{\mu_{s,k}^\star>0\}$, $P_e^\star\geq 0$ and $\{P_{k}^\star\geq 0\}$ are the optimal primal variables, $\{\varpi^\star_k\geq 0\}$ and $\{\xi_k^{\star}\geq 0\}$ are the optimal dual variables.

Due to $\{\mu_{s,k}^{\star}>0\}$ and $\{P_k^{\star}\geq 0\}$, it follows from (\ref{kkt.4}) that $\{\xi_k^{\star}> 0\}$ must be satisfied.
Then  we know from (\ref{kkt.6}) that
\begin{align}
\frac{\mu_{s,k}^\star P_{k}^\star \|\tilde{\bm{h}}_{k}\|^2}{\mu_{s,k}^\star\gamma_k+P_{k}^\star}-\sigma_k^2-\mu_{s,k}^\star\epsilon_k^2=0,\ \forall  k\in \{1,\ldots,K\},\label{p2.61}
\end{align}
and from (\ref{kkt.2}) that
\begin{align}
\mu_{s,k}^\star&=\frac{P_k^\star \|\tilde{\bm{h}}_{k}\|-P_k^\star \epsilon_k}{\gamma_k\epsilon_k},\ \forall  k\in \{1,\ldots,K\}. \label{p2.63}
\end{align}

Next, substituting (\ref{p2.63}) into (\ref{kkt.4}), there holds
\begin{align}
\xi_k^\star&=\frac{\gamma_k}{(\|\tilde{\bm{h}}_{k}\|-\epsilon_k)^2},\ \forall  k\in \{1,\ldots,K\}. \label{p2.65}
\end{align}

Lastly, substituting (\ref{p2.63}) and (\ref{p2.65}) into (\ref{p2.61}), we have
\begin{align}
P_{k}^\star=\frac{\gamma_k\sigma_k^2}{(\|\tilde{\bm{h}}_{k}\|-\epsilon_k)^2},\ \forall  k\in \{1,\ldots,K\}. \label{p2.66}
\end{align}
From the above discussions, we know that the optimal solutions of the proposed design $\mathbf{P1}$ could be derived in closed-form, which will significantly reduce the computational complexity for massive MIMO systems where the number of antennas is huge.

\section{Simulation Results}\label{Sec.5}

In this section, computer simulations are presented to
evaluate the performance of the proposed SINR-based robust beamforming algorithm for the physical layer security of BDMA massive MIMO systems.
It is assumed that all the legal users and the Eve are equipped with single antenna, where the received noises per antenna for all users are generated as independent CSCG random variables distributed with $\mathcal{CN}(0,1)$.
For simplicity, the required SINR for all users are assumed to be the same, i.e., $\gamma=\gamma_1,\ldots,=\gamma_K$ and $\gamma_e=\gamma_{e,1},\ldots,=\gamma_{e,K}$.
Define $g=\epsilon_k/ \|\tilde{\bm{h}}_{k}\|=\epsilon_e/ \|\tilde{\bm{h}}_{e}\|$ with $g \in [0,1)$.
The simulation results are averaged over 10000 Monte Carlo runs.

\begin{figure}[t]
    \centering
    \includegraphics[scale=0.6]{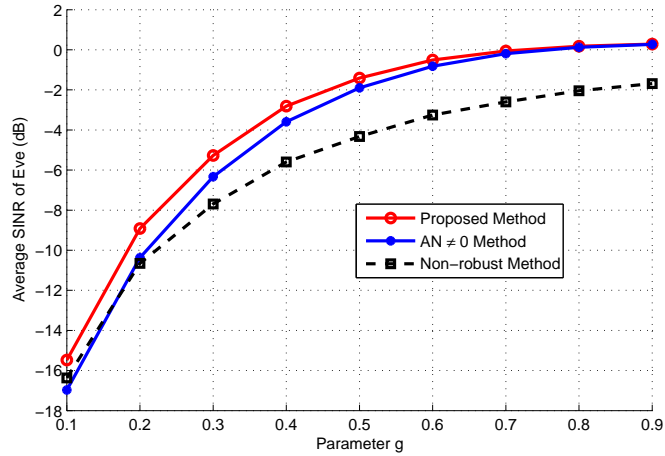}
    \caption{Average SINR of Eve versus the parameter $g$ with $K=30$ and $N=128$.}\label{fig2}
\end{figure}

In the  first example, we plot the average SINR of Eve versus the parameter $g$ with $K=30$, $N=128$ and $\gamma=10$ dB for the proposed method,
the AN $\neq 0$ method\footnote{In this method we assume that the AN beamforming power is set as $0.3\sum_{k=1}^{K}P_{k}^\star$ and the information beamforming power is set as $0.7P_{k}^\star$ for each legal user. Thus, the power consumption of the AN $\neq 0$ method is exactly the same as that of the proposed method.} and the non-robust method\footnote{Note that for the  non-robust method, the estimated channels will be directly assumed as perfect and are used for beamforming.  Thus, the optimal beamforming solutions of $\mathbf{P1}$ for the non-robust method can be easily derived as $\bm{w}_e=0$ and $\bm{s}_k=\gamma_k\sigma_k^2/\|\tilde{\bm{h}}_{k}\|$.  The details are omitted here for brevity.} in Fig.~\ref{fig2}.
Note that the theoretical results in Section~\ref{Sec.4} say that the optimal power allocation for AN beamforming should be zero,
which implies that the received SINR of Eve is smaller than $\gamma_e$ (please cf. Eq. (\ref{p1.8}) for details) without using AN.
These theoretical results can be reflected in Fig.~\ref{fig2} where it is clear that the average SINRs of Eve obtained by the proposed method, the AN $\neq 0$ method and the non-robust method are smaller than 0 dB when $g\leq 0.7$. This phenomenon implies that the constraint in  (\ref{p1.8}) will be strictly satisfied when we set $\gamma_e>0$ dB.
Based on the results of Fig.~\ref{fig2}, we could always choose a small $\gamma_e$, i.e., $0<\gamma_e<1$ to guarantee that (\ref{p1.8}) is strictly satisfied for the proposed robust design.
Moreover, it is obvious in Fig.~\ref{fig2} that the average SINRs of Eve derived by the three methods will increase with the increase of $g$.
This is mainly due to the fact that when $g$ becomes large, i.e., the channel estimation errors grow large, more power will be received at Eve.
Nevertheless, we can assume that $g$ is a small constant which is reasonable since large channel estimate errors usually lead to unacceptable degradation in performance.
It is seen from Fig.~\ref{fig2} that the average SINR of Eve derived by the proposed method is always bigger than that of the AN $\neq 0$ method and the non-robust method.
This phenomenon can be explained by the following simulation results, i.e., the proposed method will obtain a higher secret sum-rate.

\begin{figure}[t]
    \centering
    \includegraphics[scale=0.6]{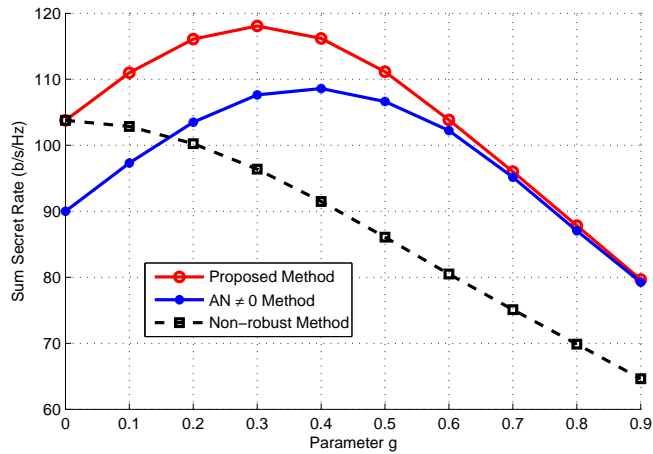}
    \caption{Secret sum-rate versus the parameter $g$ with $K=30$, $N=128$ and $\gamma=10$ dB.}\label{fig3}
\end{figure}
\begin{figure}[t]
    \centering
    \includegraphics[scale=0.6]{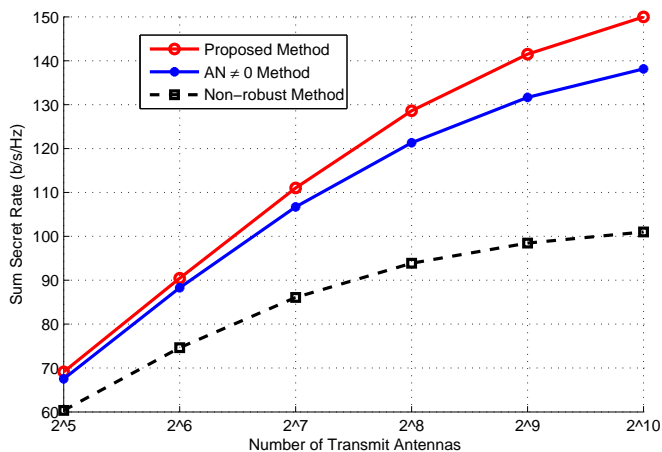}
    \caption{Secret sum-rate versus number of transmit antennas $N$ with $K=30$,  $g=0.5$ and $\gamma=10$ dB.}\label{fig4}
\end{figure}

\begin{figure}[t]
    \centering
    \includegraphics[scale=0.6]{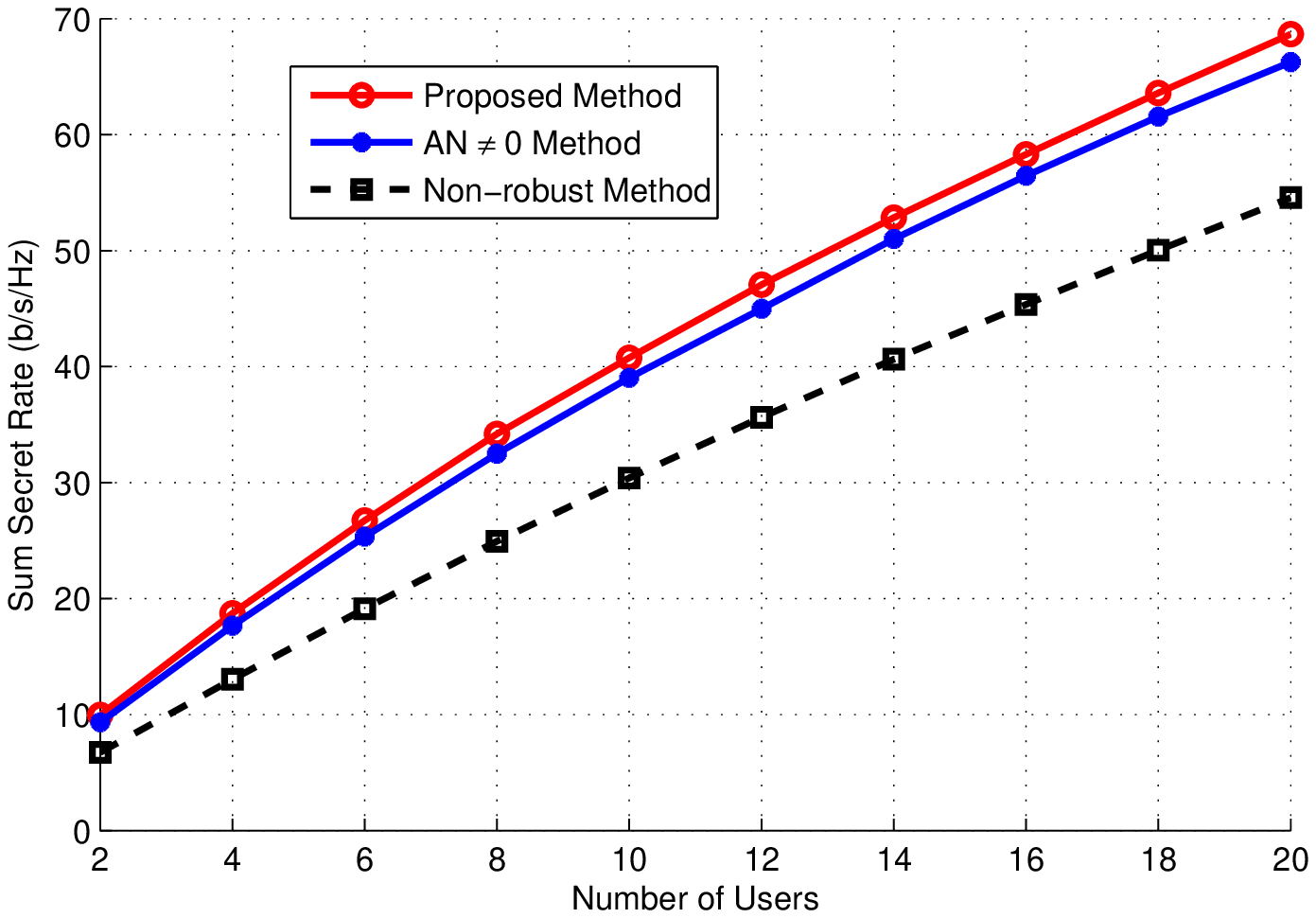}
    \caption{Secret sum-rate with different number of uses $K$ for $N=128$,  $g=0.5$ and $\gamma=10$ dB.}\label{fig5}
\end{figure}

In the second example, we plot the secret sum-rate versus parameter $g$, $N$ and $K$, in Fig.~\ref{fig3}, Fig.~\ref{fig4} and Fig.~\ref{fig5}, respectively.
Note that the secret sum-rate derived by the proposed method is always bigger than that of the AN $\neq 0$ method.
Thus, in order to optimally guarantee the the physical layer security for BDMA massive MIMO system, one should not adopt the AN beamforming.
Moreover, it is clear in Fig.~\ref{fig3} that the secret sum-rate for the non-robust method will decrease with parameter $g$ increasing.
This is mainly due to the fact that the non-robust method does not take the channel estimation errors into consideration.
Thus, with $g$ increasing, more power will be received at Eve and the secret sum-rate performance will be deteriorated.
Interestingly, the secret sum-rate for the proposed method will increase with the increase of $g$ when $g\leq 0.3$.
The reason lies in that when $g$ increases, BS will allocate more power for information beamforming (please cf. Eq. (\ref{p2.66}) for details).
While the secret sum-rate for the proposed method will decrease with the increase of $g$ when $g> 0.3$.
This phenomenon can be explained by Fig.~\ref{fig2} where it says that the average SINR of Eve will also increase with the increase of $g$.
Thus, the secret sum-rate for the proposed method will decrease when the increase of SINR of legal users is smaller than that of the Eve.
Similarly, it is seen from Fig.~\ref{fig4} and Fig.~\ref{fig5} that the secret sum-rates for the proposed method, the AN $\neq 0$ method and the non-robust method will increase with the increase of $N$ and $K$.

\begin{figure}[t]
    \centering
    \includegraphics[scale=0.6]{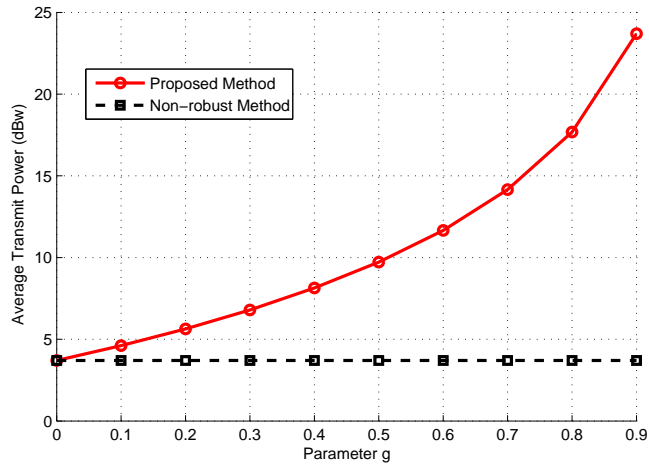}
    \caption{Average minimum transmit power versus $g$ with $K=30$, $N=128$ and $\gamma=10$ dB.}\label{fig6}
\end{figure}

\begin{figure}[t]
    \centering
    \includegraphics[scale=0.6]{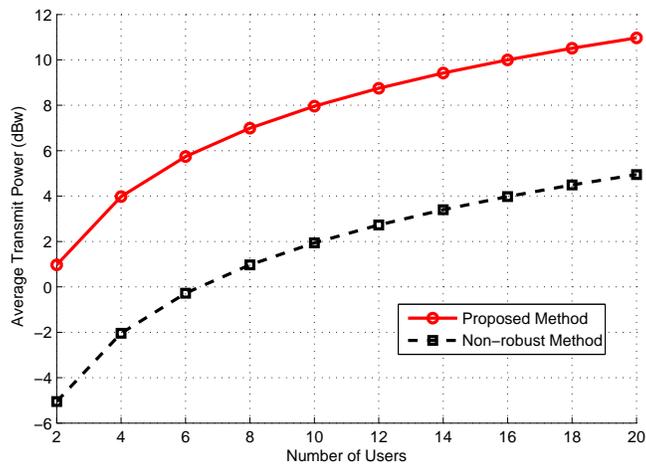}
    \caption{Average minimum transmit power versus $N$  with  $K=30$, $g=0.5$ and $\gamma=10$ dB.}\label{fig7}
\end{figure}

\begin{figure}[t]
    \centering
    \includegraphics[scale=0.6]{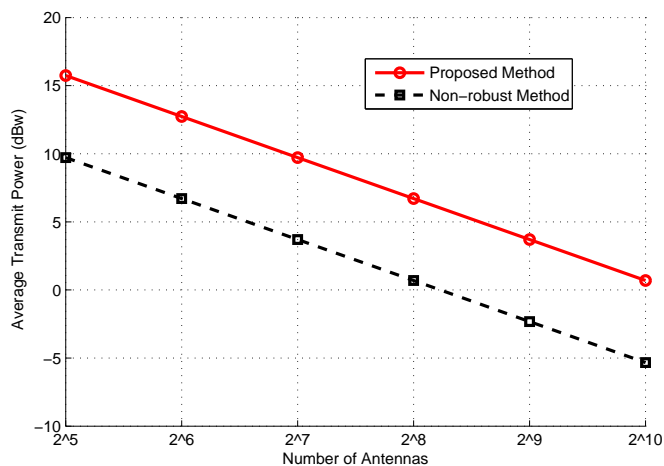}
    \caption{Average minimum transmit power versus $K$  with  $N=128$, $g=0.5$ and $\gamma=10$ dB.}\label{fig8}
\end{figure}

In the last example, we plot the average minimum transmit power versus the parameter $g$, $N$ and $K$ in Fig.~\ref{fig6}, Fig.~\ref{fig7} and Fig.~\ref{fig8}, respectively.
Note that the power consumption of the AN $\neq 0$ method is exactly the same as that of the proposed method.
Thus, the simulation results of the AN $\neq 0$ method are not contained in the last example.
It is seen from Fig.~\ref{fig6} that the transmit power for the non-robust method will not change with parameter $g$.
While the transmit power for the proposed method will increase when $g$ becomes large.
This is because when the channel estimation errors becomes large, the BS will allocate more power to eliminate the CSI uncertainty.
As a result, the proposed method can obtain a higher secret sum-rate than the non-robust method as in Fig.~\ref{fig3}, Fig.~\ref{fig4} and Fig.~\ref{fig5}.
Moreover, it is clear in Fig.~\ref{fig7} that the transmit power for both the proposed method and the non-robust method will increase with $K$ increasing,
and in Fig.~\ref{fig8} that the transmit power for both the proposed method and the non-robust method will decrease with the increase of $N$.


\section{Conclusions}\label{Sec.6}

In this paper, we  design simultaneous information and AN beamforming to guarantee the physical layer security for multiuser BDMA massive MIMO systems.
Taking the the channel estimation errors into consideration, our target is to minimize the transmit power of BS meanwhile provide the legal users and the Eve with different SINRs.
The original problem is NP-hard and optimal solutions are generally unavailable in conventional communication systems.
Nevertheless, it is strictly proved that the initial non-convex optimization can be equivalently reformulated as an SDP, where rank-one solutions are guaranteed in BDMA multiuser massive MIMO systems.
More importantly, the overall global optimal solutions for the original problem are derived in closed-form, which will greatly reduce the computational complexity. An interesting phenomenon is that the AN
is not useful for multi-user BDMA massive scenario.
Simulation results are provided to corroborate the proposed studies.

\numberwithin{equation}{section}

\appendices

\section{Proof of Proposition 1} \label{Appen:A}

\subsection {Proof of Part (a)}

Let us first show that $\{\mu_{e,k}^\star>0\}$ must hold from contradiction.
Assuming $\mu_{e,k}^\star=0$ for some $k$ that $k\in \{1,\ldots,K\}$, it follows from (\ref{p1.eqv.3}) that
\begin{align}
\bm{\Upsilon}_{e,k}=\left[ \begin{array}{ccc}
\bm{X}_{e,k}  & \bm{X}_{e,k}\tilde{\bm{h}}_e \\
\tilde{\bm{h}}_e^{\rm H}\bm{X}_{e,k}^{\rm H} & \tilde{\bm{h}}_e^{\rm H}\bm{X}_{e,k}\tilde{\bm{h}}_e+\sigma_e^2
\end{array} \right]\succeq \bm{0}\label{Appen.a.1}
\end{align}
must be satisfied.
Then it is obvious in (\ref{Appen.a.1}) that
\begin{align}
\bm{X}_{e,k}=\displaystyle\sum_{i\neq k}\bm{S}_i+\bm{W}_e-\frac{1}{\gamma_{e,k}}\bm{S}_k\succeq \bm{0},\label{Appen.a.2}
\end{align}
must be satisfied.
Moreover, from (\ref{p1.eqv.3}) and (\ref{p1.eqv.4}) , we can obtain
\begin{align}
\bm{X}_{s,k}=-\bm{X}_{e,k}-(\frac{1}{\gamma_{e,k}}-\frac{1}{\gamma_k})\bm{S}_k\preceq \bm{0},\label{Appen.a.3}
\end{align}
where the last ``$\preceq$'' holds true due to the fact that $\bm{X}_{e,k}\succeq \bm{0}$, $\bm{S}_k\succeq \bm{0}$ and $\gamma_{e,k}<\gamma_k$.

Next, we know from  (\ref{p1.eqv.2}) that
\begin{align}
\tilde{\bm{h}}_k^{\rm H}\bm{X}_{s,k}\tilde{\bm{h}}_k-\sigma_k^2-\mu_{s,k}\epsilon_k^2\geq 0,\label{Appen.a.4}
\end{align}
should be strictly satisfied. However, substituting (\ref{Appen.a.3}) into (\ref{Appen.a.4}), we will obtain
\begin{align}
0\leq \tilde{\bm{h}}_k^{\rm H}\bm{X}_{s,k}\tilde{\bm{h}}_k-\sigma_k^2-\mu_{s,k}\epsilon_k^2\leq -\sigma_k^2-\mu_{s,k}\epsilon_k^2\leq  -\sigma_k^2 ,\label{Appen.a.5}
\end{align}
where the last ``$\leq$'' is satisfied since $\mu_{s,k}\geq 0$ and $\epsilon_k^2>0$.
Due to the fact that $\sigma_k^2>0$, we know from (\ref{Appen.a.5}) can not be true.
Thus,  at the optimal point, $\mu_{e,k}^\star>0,\ \forall k\in \{1,\ldots,K\}$ must be satisfied in $\mathbf{P1\!-\!SDR}$.

\subsection {Proof of Part (b)}

Then, let us show that $\{\mu_{s,k}^\star>0\}$ must hold from contradiction.
Assuming $\mu_{s,k}^\star=0$ for some $k$ that $k\in \{1,\ldots,K\}$, it follows from (\ref{p1.eqv.4}) that
\begin{align}
\bm{\Upsilon}_{s,k}=\left[ \begin{array}{ccc}
\bm{X}_{s,k}  & \bm{X}_{s,k}\tilde{\bm{h}}_k \\
\tilde{\bm{h}}_k^{\rm H}\bm{X}_{s,k}^{\rm H} & \tilde{\bm{h}}_k^{\rm H}\bm{X}_{s,k}\tilde{\bm{h}}_k-\sigma_k^2
\end{array} \right]\succeq \bm{0},
\end{align}
must be satisfied. Left and right multiplying both sides of $\bm{\Upsilon}_{s,k}$ by $[-\tilde{\bm{h}}_k^{\rm H} \ 1]$ and $[-\tilde{\bm{h}}_k^{\rm H} \ 1]^{\rm H}$, respectively,  yields
\begin{align}
[-\tilde{\bm{h}}_k^{\rm H} \ 1]\bm{\Upsilon}_{s,k}[-\tilde{\bm{h}}_k^{\rm H} \ 1]^{\rm H}=-\sigma_k^2\geq 0,
\end{align}
which cannot be true due to $\sigma_k^2>0$. Thus, at the optimal point, $\mu_{s,k}^\star>0,\ \forall k\in \{1,\ldots,K\}$ must be satisfied in $\mathbf{P1\!-\!SDR}$.

The proof is  thus completed.


\section{Proof of Proposition 2} \label{Appen:B}

\subsection {Proof of Part (a)}

Let us show that ${\rm Rank}\left(\bm{X}_{e,k}^\star+\mu_{e,k}^\star\bm{I}\right)\geq 1$, $\forall k\in \{1,\ldots,K\}$ from contradiction.
Assume ${\rm Rank}\left(\bm{X}_{e,k}^\star+\mu_{e,k}^\star\bm{I}\right)=0$ or $\bm{X}_{e,k}^\star+\mu_{e,k}^\star\bm{I}=\bm{0}$ at the optimal point, for some $k$ that $\forall k\in \{1,\ldots,K\}$.
It is known from (\ref{p1.eqv.4}) that $\bm{X}_{e,k}^\star+\mu_{e,k}^\star\bm{I}$ can be re-expressed as
\begin{align}
\bm{X}_{e,k}^\star+\mu_{e,k}^\star\bm{I}=\mu_{e,k}^\star\bm{I}+ \displaystyle\sum_{i\neq k}\bm{S}_i^\star+\bm{W}_e^\star-\displaystyle\frac{1}{\gamma_{e,k}}\bm{S}_k^\star=\bm{0}.\label{Appen.b.1}
\end{align}
From Proposition 1, we know that $\mu_{e,k}^\star>0$ is strictly satisfied at the optimal point.
Moreover, due to the fact that $\{\bm{S}_k^\star\succeq \bm{0}\}$ and $\bm{W}_e^\star\succeq \bm{0}$, there must be
\begin{align}
{\rm Rank}\left(\mu_{e,k}^\star\bm{I}+ \displaystyle\sum_{i\neq k}\bm{S}_i^\star+\bm{W}_e^\star\right)=N, \label{Appen.b.2}
\end{align}
which says that $\mu_{e,k}^\star\bm{I}+ \displaystyle\sum_{i\neq k}\bm{S}_i^\star+\bm{W}_e^\star\succ \bm{0}$.
Then from (\ref{Appen.b.1}) and (\ref{Appen.b.2}) we know that when $\bm{X}_{e,k}^\star+\mu_{e,k}^\star\bm{I}=\bm{0}$, there must be $\bm{S}_{k}^\star\succ \bm{0}$ or ${\rm Rank}\left(\bm{S}_k^\star\right) =N$.

Define $N\times N$ matrix $\bm{Q}$ as
\begin{align}
\bm{Q}=\left[\frac{\tilde{\bm{h}}_e}{\|\tilde{\bm{h}}_e\|},\frac{\tilde{\bm{h}}_1}{\|\tilde{\bm{h}}_1\|},\ldots,\frac{\tilde{\bm{h}}_K}{\|\tilde{\bm{h}}_K\|},\bm{\tau}_{1},\ldots,\bm{\tau}_{N-K-1}\right],\label{Appen.b.3}
\end{align}
where $\bm{\tau}_{i}^{N\times 1}, \forall i\in\{1,\ldots,N-K-1\}$ is a unit vector, i.e., $\|\bm{\tau}_{i}\|=1$ which is constrained to satisfy the following properties:
\begin{enumerate}
\item [(a)] $\bm{\tau}_{i}\perp \bm{\tau}_{j},\forall i\neq j$;
\item [(b)] $\bm{\tau}_{i}\perp \tilde{\bm{h}}_e, \forall i\in\{1,\ldots,N-K-1\}$;
\item [(c)] $\bm{\tau}_{i}\perp \tilde{\bm{h}}_k, \forall i\in\{1,\ldots,N-K-1\}, \forall k\in\{1,\ldots,K\}$.
\end{enumerate}
Thanks to (\ref{model.3}) under BDMA massive MIMO scheme, we know that the defined matrix $\bm{Q}$ is consisted of $N$ orthogonal bases.
Thus, $\bm{Q}$ is invertible, i.e., $\bm{Q}^{-1}$ exists. Consequently, there must be
\begin{align}
&{\rm Rank}\left(\bm{Q}\bm{W}_e^\star\right)={\rm Rank}\left(\bm{W}_e^\star\right),\label{Appen.b.4}\\
&{\rm Rank}\left(\bm{Q}\bm{S}_k^\star\right)={\rm Rank}\left(\bm{S}_k^\star\right), \forall k\in\{1,\ldots,K\}. \label{Appen.b.5}
\end{align}
From (\ref{Appen.b.3}), (\ref{Appen.b.4})  and (\ref{Appen.b.5}), we know that $\bm{W}_{e}^{\star}$ and $\bm{S}_{k}^{\star}, \forall k\in\{1,\ldots,K\}$ can be further expressed as
\begin{align}
\bm{W}_{e}^{\star}&=P_{w_e,h_e}^\star\frac{\tilde{\bm{h}}_e\tilde{\bm{h}}_e^{\rm H}}{\|\tilde{\bm{h}}_e\|^2}+\sum_{i=1}^{K}P_{w_e,h_i}^\star\frac{\tilde{\bm{h}}_i\tilde{\bm{h}}_i^{\rm H}}{\|\tilde{\bm{h}}_i\|^2}+\sum_{i=1}^{N-K-1}P_{w_e,\tau_i}^\star\bm{\tau}_i\bm{\tau}_i^{\rm H},\label{Appen.b.6}\\
\bm{S}_{k}^{\star}&=P_{s_k,h_e}^\star\frac{\tilde{\bm{h}}_e\tilde{\bm{h}}_e^{\rm H}}{\|\tilde{\bm{h}}_e\|^2}+\sum_{i=1}^{K}P_{s_k,h_i}^\star\frac{\tilde{\bm{h}}_i\tilde{\bm{h}}_i^{\rm H}}{\|\tilde{\bm{h}}_i\|^2}+\sum_{i=1}^{N-K-1}P_{s_k,\tau_i}^\star\bm{\tau}_i\bm{\tau}_i^{\rm H},\label{Appen.b.7}
\end{align}
where $P_{w_e,h_e}^\star\geq 0$, $\{P_{w_e,h_i}^\star\geq 0\}$, $\{P_{w_e,\tau_i}^\star\geq 0\}$, $P_{s_k,h_e}^\star\geq 0$, $\{P_{s_k,h_i}^\star\geq 0\}$ and  $\{P_{s_k,\tau_i}^\star\geq 0\}$. Note that if ${\rm Rank}\left(\bm{S}_k^\star\right) =N$, then there must be $P_{s_k,h_e}^\star> 0$, $\{P_{s_k,h_i}^\star> 0\}$ and  $\{P_{s_k,\tau_i}^\star> 0\}$.
Under the previous assumption, there is at least one $\bm{S}_k^\star$ that satisfies ${\rm Rank}\left(\bm{S}_k^\star\right) =N$, where $P_{s_k,h_e}^\star> 0$, $\{P_{s_k,h_i}^\star> 0\}$ and  $\{P_{s_k,\tau_i}^\star> 0\}$ must hold.

Then, letting $\{P_{w_e, \tau_i}^\star=0\}, \{P_{s_k, \tau_i}^\star=0\}, \forall i\in \{1,\ldots,N-K-1\}, \forall k\in \{1,\ldots,K\}$, we can construct the following new solutions
\begin{align}
\left\{ \begin{array}{ll}
\bm{W}_{e}^{*}=\displaystyle P_{w_e,h_e}^\star\frac{\tilde{\bm{h}}_e\tilde{\bm{h}}_e^{\rm H}}{\|\tilde{\bm{h}}_e\|^2}+\sum_{i=1}^{K}P_{w_e,h_i}^\star\frac{\tilde{\bm{h}}_i\tilde{\bm{h}}_i^{\rm H}}{\|\tilde{\bm{h}}_i\|^2},\\
\bm{S}_{k}^{*}=\displaystyle P_{s_k,h_e}^\star\frac{\tilde{\bm{h}}_e\tilde{\bm{h}}_e^{\rm H}}{\|\tilde{\bm{h}}_e\|^2}+\sum_{i=1}^{K}P_{s_k,h_i}^\star\frac{\tilde{\bm{h}}_i\tilde{\bm{h}}_i^{\rm H}}{\|\tilde{\bm{h}}_i\|^2}.
\end{array} \right.\label{Appen.b.8}
\end{align}
Substituting (\ref{Appen.b.8}) into $\mathbf{P1\!-\!SDR}$ and using the properties of (\ref{Appen.b.3}), we obtain
\begin{align}
&  {\rm Tr}(\bm{W}_e^\star)+\sum_{k=1}^{K}{\rm Tr}(\bm{S}_k^\star)> {\rm Tr}(\bm{W}_e^*)+\sum_{k=1}^{K}{\rm Tr}(\bm{S}_k^*)\label{Appen.b.7a}\\
  &   \left[ \begin{array}{ccc}
\bm{X}_{e,k}^*+\mu_{e,k}^{\star}\bm{I} & \bm{X}_{e,k}^*\tilde{\bm{h}}_e \\
\tilde{\bm{h}}_e^{\rm H}\bm{X}_{e,k}^{*\rm H} & \tilde{\bm{h}}_e^{\rm H}\bm{X}_{e,k}^*\tilde{\bm{h}}_e+\sigma_e^2-\mu_{e,k}^\star\epsilon_e^2
\end{array} \right]\succeq \bm{0},\label{Appen.b.8a}\\
&    \left[ \begin{array}{ccc}
\bm{X}_{s,k}^*+\mu_{s,k}^\star\bm{I} & \bm{X}_{s,k}^*\tilde{\bm{h}}_k \\
\tilde{\bm{h}}_k^{\rm H}\bm{X}_{s,k}^{*\rm H} & \tilde{\bm{h}}_k^{\rm H}\bm{X}_{s,k}^*\tilde{\bm{h}}_k-\sigma_k^2-\mu_{s,k}^\star\epsilon_k^2
\end{array} \right]\succeq \bm{0},\label{Appen.b.9a}\\
& \bm{X}_{e,k}^*=\displaystyle\sum_{i\neq k}\bm{S}_i^*+\bm{W}_e^*-\displaystyle\frac{1}{\gamma_{e,k}}\bm{S}_k^*, \label{Appen.b.10a}\\
&  \bm{X}_{s,k}^*=\displaystyle\frac{1}{\gamma_k}\bm{S}_k^*-\displaystyle\sum_{i\neq k}\bm{S}_i^*-\bm{W}_e^*, \label{Appen.b.11a}\\
&  \mu_{e,k}^\star\geq 0,\quad \mu_{s,k}^\star\geq 0,\quad \bm{W}_e^*\succeq \bm{0},\quad \bm{S}_k^*\succeq \bm{0},\quad k=1,2,\ldots,K.   \label{Appen.b.12a}
\end{align}
From (\ref{Appen.b.8a})$\sim$(\ref{Appen.b.12a}) we know that $\bm{W}_e^*$ and $\{\bm{S}_k^*\}$ satisfy all constraints of $\mathbf{P1\!-\!SDR}$.
From (\ref{Appen.b.7a}) we know that $\bm{W}_e^*$ and $\{\bm{S}_k^*\}$ will always provide smaller objective value.
Consequently, $\bm{W}_e^*$ and $\{\bm{S}_k^*\}$ are better solutions than $\bm{W}_e^\star$ and $\{\bm{S}_k^\star\}$, which contradicts with our first place assumption.
Thus, there must be ${\rm Rank}\left(\bm{S}_k^\star\right) <N$ and ${\rm Rank}\left(\bm{X}_{e,k}^\star+\mu_{e,k}^\star\bm{I}\right)\geq 1$, $\forall k\in \{1,\ldots,K\}$.

\subsection {Proof of Part (b)}

Let us show that ${\rm Rank}\left(\bm{X}_{s,k}^\star+\mu_{s,k}^\star\bm{I}\right)\geq 1$, $\forall k\in \{1,\ldots,K\}$ from contradiction.
Assume ${\rm Rank}\left(\bm{X}_{s,k}^\star+\mu_{s,k}^\star\bm{I}\right)=0$ or $\bm{X}_{s,k}^\star+\mu_{s,k}^\star\bm{I}=\bm{0}$ at the optimal point, for some $k$ that $\forall k\in \{1,\ldots,K\}$. Due to the fact that $\mu_{s,k}^\star>0, \forall k\in \{1,\ldots,K\}$ (please cf. Proposition 1 for details), there must be
\begin{align}
\bm{X}_k^\star=-\mu_{s,k}^\star\bm{I}\preceq \bm{0}.\label{Appen.b.15}
\end{align}
Then substituting (\ref{Appen.b.15}) into  (\ref{p1.eqv.2}), it is easily known that
\begin{align}
 \left[ \begin{array}{ccc}
\bm{0} & \bm{X}_{s,k}^\star\tilde{\bm{h}}_k \\
\tilde{\bm{h}}_k^{\rm H}\bm{X}_{s,k}^{\star\rm H} & \tilde{\bm{h}}_k^{\rm H}(-\mu_{s,k}^\star\bm{I})\tilde{\bm{h}}_k-\sigma_k^2-\mu_{s,k}^\star\epsilon_k^2
\end{array} \right]\succeq \bm{0},
\end{align}
must be strictly satisfied.
which cannot be true due to the fact that $\tilde{\bm{h}}_k^{\rm H}(-\mu_{s,k}^\star\bm{I})\tilde{\bm{h}}_k-\sigma_k^2-\mu_{s,k}^\star\epsilon_k^2< 0$.
Thus there must be ${\rm Rank}\left(\bm{X}_{s,k}^\star+\mu_{s,k}^\star\bm{I}\right)\geq 1$, $\forall k\in \{1,\ldots,K\}$.

The proof of Proposition 2 is completed.

\section{Proof of Theorem 1} \label{Appen:C}

The following lemma is useful in the proof of Theorem 1.

\emph{Lemma 3  (Theory of Majorization \cite{Marshall1979}):} Let $\bm{A}$ and $\bm{B}$ be two $N\times N$ positive semi-definite matrices with eigenvalues $\alpha_1\geq\ldots\geq\alpha_N$ and $\beta_1\geq\ldots\geq\beta_N$, respectively. Then there holds
\begin{align}
\displaystyle\sum_{i=1}^{N}\alpha_i\beta_{N-i+1}\leq {\rm Tr}(\bm{A}\bm{B})\leq \displaystyle\sum_{i=1}^{N}\alpha_i\beta_i.\label{Appen.C.1}
\end{align}

Firstly, it    follows from (\ref{Appen.b.8}) that at the optimal point, $\bm{W}_e^{\star}$ and $\{\bm{S}_k^{\star}\}$ can be expressed as
\begin{align}
\left\{ \begin{array}{ll}
\bm{W}_{e}^{\star}=\displaystyle P_{w_e,h_e}^\star\frac{\tilde{\bm{h}}_e\tilde{\bm{h}}_e^{\rm H}}{\|\tilde{\bm{h}}_e\|^2}+\sum_{i=1}^{K}P_{w_e,h_i}^\star\frac{\tilde{\bm{h}}_i\tilde{\bm{h}}_i^{\rm H}}{\|\tilde{\bm{h}}_i\|^2},\\
\bm{S}_{k}^{\star}=\displaystyle P_{s_k,h_e}^\star\frac{\tilde{\bm{h}}_e\tilde{\bm{h}}_e^{\rm H}}{\|\tilde{\bm{h}}_e\|^2}+\sum_{i=1}^{K}P_{s_k,h_i}^\star\frac{\tilde{\bm{h}}_i\tilde{\bm{h}}_i^{\rm H}}{\|\tilde{\bm{h}}_i\|^2},
\end{array} \right.\label{Appen.C.2}
\end{align}
where $P_{w_e,h_e}^\star\geq 0$ and $\{P_{w_e,h_i}^\star\geq 0\}$ are the coefficients of $\bm{W}_{e}^{\star}$, while $P_{s_k,h_e}^\star\geq 0$ and  $\{P_{s_k,h_i}^\star\geq 0\}$ are the coefficients of $\bm{S}_{k}^{\star}$.
Then using (\ref{Appen.C.2}),   $\bm{X}_{e,k}^\star$ and $\bm{X}_{s,k}^\star$ can be expressed as
\begin{align}
&\bm{X}_{e,k}^{\star}=\displaystyle\sum_{i\neq k}\bm{S}_i^{\star}+\bm{W}_e^{\star}-\frac{1}{\gamma_{e,k}}\bm{S}_k^{\star}  \label{Appen.C.3} \\
&=\left(\sum_{i\neq k}P_{s_i,h_e}^\star+P_{w_e,h_e}^\star-\frac{1}{\gamma_{e,k}}P_{s_k,h_e}^\star\right)\frac{\tilde{\bm{h}}_e\tilde{\bm{h}}_e^{\rm H}}{\|\tilde{\bm{h}}_e\|^2}
+\sum_{j=1}^K\left(\sum_{i\neq k}P_{s_i,h_j}^\star+P_{w_e,h_j}^\star-\frac{1}{\gamma_{e,k}}P_{s_k,h_j}^\star\right)\frac{\tilde{\bm{h}}_j\tilde{\bm{h}}_j^{\rm H}}{\|\tilde{\bm{h}}_j\|^2}, \nonumber\\
&\bm{X}_{s,k}^{\star}=\displaystyle\frac{1}{\gamma_k}\bm{S}_k^{\star}-\sum_{i\neq k}\bm{S}_i^{\star}-\bm{W}_e^{\star}  \label{Appen.C.4} \\
&=\left(\frac{1}{\gamma_k}P_{s_k,h_e}^\star-\sum_{i\neq k}P_{s_i,h_e}^\star-P_{w_e,h_e}^\star\right)\frac{\tilde{\bm{h}}_e\tilde{\bm{h}}_e^{\rm H}}{\|\tilde{\bm{h}}_e\|^2}
+\sum_{j=1}^K\left(\frac{1}{\gamma_k}P_{s_k,h_j}^\star-\sum_{i\neq k}P_{s_i,h_j}^\star-P_{w_e,h_j}^\star\right)\frac{\tilde{\bm{h}}_j\tilde{\bm{h}}_j^{\rm H}}{\|\tilde{\bm{h}}_j\|^2}. \nonumber
\end{align}
With the BDMA massive MIMO regime, the estimated channel vectors satisfy $\tilde{\bm{h}}_i\perp\tilde{\bm{h}}_k\perp\tilde{\bm{h}}_e,\ \forall i\neq k$.
Thus, we know from (\ref{Appen.C.2})$\sim$(\ref{Appen.C.4}) that $\tilde{\bm{h}}_{e}/\|\tilde{\bm{h}}_{e}\|, \tilde{\bm{h}}_{1}/\|\tilde{\bm{h}}_{1}\|,\ldots, \tilde{\bm{h}}_{K}/\|\tilde{\bm{h}}_{K}\|$ $\in$ $\{\bm{u}_{e,k,i}\}$ and $\tilde{\bm{h}}_{e}/\|\tilde{\bm{h}}_{e}\|, \tilde{\bm{h}}_{1}/\|\tilde{\bm{h}}_{1}\|,\ldots, \tilde{\bm{h}}_{K}/\|\tilde{\bm{h}}_{K}\|$ $\in$ $\{\bm{u}_{s,k,i}\}$, where $\bm{u}_{e,k,i}$ and $\bm{u}_{s,k,i}$ are the $i$th column of $\bm{U}_{e,k}$ and $\bm{U}_{s,k}$ respectively. Based on the above discussions, we know from  (\ref{3.2.42}) that:
\begin{enumerate}
\item [(a)]  If $\tilde{\bm{h}}_{e}/\|\tilde{\bm{h}}_{e}\|$ lies in the null space of $\bm{X}_{e,k}^{\star}+\mu_{e,k}^\star\bm{I}$, i.e.,
$\tilde{\bm{h}}_{e}/\|\tilde{\bm{h}}_{e}\|=\bm{u}_{e,k,i}$ and $l_{e,k}+1\leq i \leq N$, there holds
    \begin{align}
    \sum_{i\neq k}P_{s_i,h_e}^\star+P_{w_e,h_e}^\star-\frac{1}{\gamma_{e,k}}P_{s_k,h_e}^\star+\mu_{e,k}^\star=0;\label{Appen.C.5}
    \end{align}
\item [(b)]  If $\tilde{\bm{h}}_{e}/\|\tilde{\bm{h}}_{e}\|$ lies in the range space of $\bm{X}_{e,k}^{\star}+\mu_{e,k}^\star\bm{I}$, i.e., $\tilde{\bm{h}}_{e}/\|\tilde{\bm{h}}_{e}\|=\bm{u}_{e,k,i}$ and $1\leq i \leq l_{e,k}$, there holds
    \begin{align}
    \sum_{i\neq k}P_{s_i,h_e}^\star+P_{w_e,h_e}^\star-\frac{1}{\gamma_{e,k}}P_{s_k,h_e}^\star+\mu_{e,k}^\star>0.\label{Appen.C.6}
    \end{align}
\end{enumerate}
Similarly, we know from  (\ref{3.2.43}) that
\begin{enumerate}
\item [(a)] If $\tilde{\bm{h}}_{k}/\|\tilde{\bm{h}}_{k}\|$ lies in the null space of $\bm{X}_{s,k}^{\star}+\mu_{s,k}^\star\bm{I}$ i.e., $\tilde{\bm{h}}_{k}/\|\tilde{\bm{h}}_{k}\|=\bm{u}_{s,k,i}$ and $l_{s,k}+1\leq i \leq N$, there holds
    \begin{align}
    \frac{1}{\gamma_k}P_{s_k,h_k}^\star-\sum_{i\neq k}P_{s_i,h_k}^\star-P_{w_e,h_k}^\star+\mu_{s,k}^\star=0;\label{Appen.C.7}
    \end{align}
\item [(b)] If $\tilde{\bm{h}}_{k}/\|\tilde{\bm{h}}_{k}\|$ lies in the range space of $\bm{X}_{s,k}^{\star}+\mu_{s,k}^\star\bm{I}$, i.e., $\tilde{\bm{h}}_{k}/\|\tilde{\bm{h}}_{k}\|=\bm{u}_{s,k,i}$ and $1\leq i \leq l_{s,k}$, there holds
    \begin{align}
    \frac{1}{\gamma_k}P_{s_k,h_k}^\star-\sum_{i\neq k}P_{s_i,h_k}^\star-P_{w_e,h_k}^\star+\mu_{s,k}^\star>0.\label{Appen.C.8}
    \end{align}
\end{enumerate}

Next, let the EVD of $\tilde{\bm{H}}_e$ be $\tilde{\bm{H}}_e=\bm{U}_{\tilde{h},e}\bm{\Lambda}_{\tilde{h},e}\bm{U}_{\tilde{h},e}^{\rm H}$, where $\bm{\Lambda}_{\tilde{h},e}$ is diagonal with the eigenvalues $\lambda_{e,1}\geq\ldots\geq \lambda_{e,N}$. Since  $\tilde{\bm{H}}_e=\tilde{\bm{h}}_e\tilde{\bm{h}}_e^{\rm H}$ has only one nonzero eigenvalue,
there must be $\lambda_{e,1}>0$, $\lambda_{e,2}=\ldots =\lambda_{e,N}=0$  and $\bm{u}_{\tilde{h},e,1}=\tilde{\bm{h}}_{e}/\|\tilde{\bm{h}}_{e}\|$, where $\bm{u}_{\tilde{h},e,1}$ is the first column of $\bm{U}_{\tilde{h},e}$.
Then according to (\ref{3.2.42}) and Lemma 3, we know that (\ref{3.2.44}) can be further reformulated as
\begin{align}
&0\leq {\rm Tr}\left\{\left(\tilde{\bm{H}}_e\bm{X}_{e,k}\right)\left[\bm{I} -\left(\bm{X}_{e,k}+\mu_{e,k}\bm{I}\right)^{\dag}\bm{X}_{e,k}\right]\right\}+\sigma_e^2-\mu_{e,k}\epsilon_e^2\nonumber\\
=&  {\rm Tr}\left\{\left(\tilde{\bm{H}}_{e,k}\bm{X}_{e,k}\right)\left[\bm{I} -\left(\bm{X}_{e,k}+\mu_{e,k}\bm{I}\right)^{\dag}\left(\bm{X}_{e,k}+\mu_{e,k}\bm{I}-\mu_{e,k}\bm{I}\right)\right]\right\}
+\sigma_e^2-\mu_{e,k}\epsilon_e^2 \nonumber\\
=& \mu_{e,k}{\rm Tr}\left[\tilde{\bm{H}}_e\bm{X}_{e,k}\left(\bm{X}_{e,k}+\mu_{e,k}\bm{I}\right)^{\dag}\right]+\sigma_e^2-\mu_{e,k}\epsilon_e^2\nonumber\\
=& \mu_{e,k}{\rm Tr}\left[\bm{\Lambda}_{e,k}\left[ \begin{array}{ccc}
\bm{\Sigma}^{-1}_{e,k,+} & \bm{0} \\
\bm{0} & \bm{0}
\end{array} \right]\bm{U}_{e,k}^{\rm H}\bm{U}_{\tilde{h},e}\bm{\Lambda}_{\tilde{h},e}\bm{U}_{\tilde{h},e}^{\rm H}\bm{U}_{e,k}\right]+\sigma_e^2-\mu_{e,k}\epsilon_e^2\nonumber\\
=&\frac{\mu_{e,k} q_{e,k,i} \lambda_{e,1}}{\mu_{e,k}+q_{e,k,i}}+\sigma_e^2-\mu_{e,k}\epsilon_e^2\leq  \frac{\mu_{e,k} q_{e,k,1} \lambda_{e,1}}{\mu_{e,k}+q_{e,k,1}}+\sigma_e^2-\mu_{e,k}\epsilon_e^2,\label{appen.c.2}
\end{align}
where $q_{e,k,i}$ is the $i$th eigenvalue of $\bm{X}_{e,k}$ whose corresponding eigenvector is $\bm{u}_{e,k,i}=\bm{u}_{\tilde{h},e,1}=\tilde{\bm{h}}_{e}/\|\tilde{\bm{h}}_{e}\|$, and $q_{e,k,1}$ is the maximum eigenvalue of $\bm{X}_{e,k}$.
Note that in (\ref{appen.c.2}), the last ``$\leq$'' holds with ``$=$'' when $\bm{u}_{e,k,1}=\bm{u}_{\tilde{h},e,1}=\tilde{\bm{h}}_{e}/\|\tilde{\bm{h}}_{e}\|$.

Similarly, let the EVD of $\tilde{\bm{H}}_k$ be $\tilde{\bm{H}}_k=\bm{U}_{\tilde{h},k}\bm{\Lambda}_{\tilde{h},k}\bm{U}_{\tilde{h},k}^{\rm H}$ where $\bm{\Lambda}_{\tilde{h},k}$ is diagonal with the single non-zero eigenvalue $\lambda_{k,1}>0$ and the first column of $\bm{U}_{\tilde{h},k}$ is $\bm{u}_{\tilde{h},k,1}=\tilde{\bm{h}}_{k}/\|\tilde{\bm{h}}_{k}\|$.
According to (\ref{3.2.43}) and Lemma A, we know that (\ref{3.2.45}) can be reformulated as (the details are omitted here for brevity)
\begin{align}
&{\rm Tr}\left\{\left(\tilde{\bm{H}}_k\bm{X}_{s,k}\right)\left[\bm{I} -\left(\bm{X}_{s,k}+\mu_{s,k}\bm{I}\right)^{\dag}\bm{X}_{s,k}\right]\right\}-\sigma_k^2-\mu_{s,k}\epsilon_k^2\nonumber\\
=& \frac{\mu_{s,k} q_{s,k,i} \lambda_{k,1}}{\mu_{s,k}+q_{s,k,i}}-\sigma_k^2-\mu_{s,k}\epsilon_k^2\leq \frac{\mu_{s,k} q_{s,k,1} \lambda_{k,1}}{\mu_{s,k}+q_{s,k,1}}-\sigma_k^2-\mu_{s,k}\epsilon_k^2,\label{appen.c.4}
\end{align}
where $q_{s,k,i}$ is the $i$th eigenvalue of $\bm{X}_{s,k}$ whose corresponding eigenvector is $\bm{u}_{s,k,i}=\bm{u}_{\tilde{h},k,1}=\tilde{\bm{h}}_{k}/\|\tilde{\bm{h}}_{k}\|$, and $q_{s,k,1}$ is the maximum eigenvalue of $\bm{X}_{s,k}$.
Note that in (\ref{appen.c.4}), the last ``$\leq$'' holds with ``$=$'' when $\bm{u}_{s,k,1}=\bm{u}_{\tilde{h},k,1}=\tilde{\bm{h}}_{k}/\|\tilde{\bm{h}}_{k}\|$, where $\bm{u}_{s,k,1}$, $\bm{u}_{\tilde{h},k,1}$ are the first columns of $\bm{U}_{s,k}$ and $\bm{U}_{\tilde{h},k}$  respectively.

Lastly, let us show that ${\rm Rank}\left(\bm{W}_e^\star\right)\leq 1$ and ${\rm Rank}\left(\bm{S}_k^\star\right)\leq 1, \ \forall k\in \{1,\ldots,K\}$ from contradiction.
Assume $\{\mu_{e,k}^{\star}\}$, $\{\mu_{s,k}^{\star}\}$, $\bm{W}_e^\star$ and $\{\bm{S}_k^\star\}$ are the optimal solutions of $\mathbf{P1\!-\!SDR\!-\!EQV}$,
where ${\rm Rank}\left(\bm{W}_e^\star\right)=C_e>1$ or ${\rm Rank}\left(\bm{S}_k^\star\right)=C_k>1$.
Then we provide the following new solutions
\begin{align}
\bm{W}_e^*=\displaystyle \left(P_{w_e,h_e}^\star+\sum_{i=1}^K P_{s_i,h_e}^\star\right)\frac{\tilde{\bm{h}}_e\tilde{\bm{h}}_e^{\rm H}}{\|\tilde{\bm{h}}_e\|^2},\quad
\bm{S}_k^*=\displaystyle \left(P_{s_k,h_k}^\star-\sum_{i\neq k}P_{s_i,h_k}^\star-P_{w_e,h_k}^\star\right)\frac{\tilde{\bm{h}}_k\tilde{\bm{h}}_k^{\rm H}}{\|\tilde{\bm{h}}_k\|^2},\label{Appen.C.12}
\end{align}
which satisfies ${\rm Rank}\left(\bm{W}_e^*\right)\leq 1$ and ${\rm Rank}\left(\bm{S}_k^*\right)\leq 1,\ \forall k\in \{1,\ldots,K\}$.
Then it is obvious that $\bm{X}_{e,k}^*$ and $\bm{X}_{s,k}^*$ can be expressed as
\begin{align}
&\bm{X}_{e,k}^*=\displaystyle\sum_{i\neq k}\bm{S}_i^*+\bm{W}_e^*-\frac{1}{\gamma_{e,k}}\bm{S}_k^*
=\sum_{i\neq k}\left(P_{s_i,h_i}^\star-\sum_{j\neq i}P_{s_j,h_i}^\star-P_{w_e,h_i}^\star\right)\frac{\tilde{\bm{h}}_i\tilde{\bm{h}}_i^{\rm H}}{\|\tilde{\bm{h}}_i\|^2}\nonumber\\
&+\left(P_{w_e,h_e}^\star+\sum_{i=1}^K P_{s_i,h_e}^\star\right)\frac{\tilde{\bm{h}}_e\tilde{\bm{h}}_e^{\rm H}}{\|\tilde{\bm{h}}_e\|^2}
-\frac{1}{\gamma_{e,k}}\left(P_{s_k,h_k}^\star-\sum_{i\neq k}P_{s_i,h_k}^\star-P_{w_e,h_k}^\star\right)\frac{\tilde{\bm{h}}_k\tilde{\bm{h}}_k^{\rm H}}{\|\tilde{\bm{h}}_k\|^2},\label{Appen.C.10}\\
&\bm{X}_{s,k}^*=\displaystyle\frac{1}{\gamma_k}\bm{S}_k^*-\sum_{i\neq k}\bm{S}_i^*-\bm{W}_e^*=\frac{1}{\gamma_{k}}\left(P_{s_k,h_k}^\star-\sum_{i\neq k}P_{s_i,h_k}^\star-P_{w_e,h_k}^\star\right)\frac{\tilde{\bm{h}}_k\tilde{\bm{h}}_k^{\rm H}}{\|\tilde{\bm{h}}_k\|^2}\nonumber\\
&-\sum_{i\neq k}\left(P_{s_i,h_i}^\star-\sum_{j\neq i}P_{s_j,h_i}^\star-P_{w_e,h_i}^\star\right)\frac{\tilde{\bm{h}}_i\tilde{\bm{h}}_i^{\rm H}}{\|\tilde{\bm{h}}_i\|^2}-\left(P_{w_e,h_e}^\star+\sum_{i=1}^K P_{s_i,h_e}^\star\right)\frac{\tilde{\bm{h}}_e\tilde{\bm{h}}_e^{\rm H}}{\|\tilde{\bm{h}}_e\|^2},\label{Appen.C.11}
\end{align}
respectively. From (\ref{Appen.C.3}) and (\ref{Appen.C.10}), we know
\begin{align}
\sum_{i\neq k}P_{s_i,h_e}^\star+P_{w_e,h_e}^\star-\frac{1}{\gamma_{e,k}}P_{s_k,h_e}^\star\leq P_{w_e,h_e}^\star+\sum_{i=1}^K P_{s_i,h_e}^\star,
\end{align}
which implies that $\bm{W}_e^*$ and $\{\bm{S}_k^*\}$ will not violate the constraints in (\ref{p1.sdr.eqv.2}) of $\mathbf{P1\!-\!SDR\!-\!EQV}$ (please cf. Eq. (\ref{Appen.C.5}) and Eq. (\ref{Appen.C.6}) for details). From (\ref{Appen.C.4}) and (\ref{Appen.C.11}), we obtain
\begin{align}
\frac{1}{\gamma_k}P_{s_k,h_k}^\star-\sum_{i\neq k}P_{s_i,h_k}^\star-P_{w_e,h_k}^\star\leq \frac{1}{\gamma_k}\left(P_{s_k,h_k}^\star-\sum_{i\neq k}P_{s_i,h_k}^\star-P_{w_e,h_k}^\star\right),
\end{align}
which implies that $\bm{W}_e^*$ and $\{\bm{S}_k^*\}$ will not violate the constraints in (\ref{p1.sdr.eqv.4}) of $\mathbf{P1\!-\!SDR\!-\!EQV}$ (please cf. Eq. (\ref{Appen.C.7}) and Eq. (\ref{Appen.C.8}) for details).
Moreover, it is easily known from (\ref{Appen.C.3}), (\ref{Appen.C.4}), (\ref{Appen.C.10}) and (\ref{Appen.C.11}) that if $\mu_{s,k}^\star-\left(P_{w_e,h_e}^\star+\sum_{i=1}^K P_{s_i,h_e}^\star\right)\geq 0$, there must hold $\bm{X}_{e,k}^*+\mu_{e,k}^\star\bm{I}\succeq \bm{0}$ and $\bm{X}_{s,k}^*+\mu_{s,k}^\star\bm{I}\succeq \bm{0}$.
Interestingly, it is observed from (\ref{appen.c.2}) that we can always let $q_{e,k,1}=P_{w_e,h_e}^\star+\sum_{i=1}^K P_{s_i,h_e}^\star=0$ and $\mu_{e,k}=\sigma_e^2/ \epsilon_e^2$, which will not violate (\ref{appen.c.2}). A more rigorous proof can be found in Theorem 2.

As a result, substituting $\bm{W}_e^*$ and $\{\bm{S}_k^*\}$ into $\mathbf{P1\!-\!SDR\!-\!EQV}$, we obtain from (\ref{p1.sdr.eqv.1}) that
\begin{align}
  & {\rm Tr}(\bm{W}_e^*)\!+\!\sum_{k=1}^{K}{\rm Tr}(\bm{S}_k^*)\!=\!P_{w_e,h_e}^\star+\sum_{i=1}^K P_{s_i,h_e}^\star+\sum_{k=1}^K \left(P_{s_k,h_k}^\star-\sum_{i\neq k}P_{s_i,h_k}^\star-P_{w_e,h_k}^\star\right)\nonumber\\
    \leq & P_{w_e,h_e}^\star+\sum_{i=1}^K P_{s_i,h_e}^\star+\sum_{k=1}^K P_{s_k,h_k}^\star+\sum_{j=1}^K\sum_{i=1}^KP_{s_i,h_j}^\star\!=\! {\rm Tr}(\bm{W}_e^\star)\!+\!\sum_{k=1}^{K}{\rm Tr}(\bm{S}_k^\star); \label{Appen.C.17}
\end{align}
from (\ref{p1.sdr.eqv.2a}) and (\ref{p1.sdr.eqv.2}) that
\begin{align}
\bm{X}_{e,k}^*+\mu_{e,k}^\star\bm{I}\succeq \bm{0},\quad \bm{X}_{s,k}^*+\mu_{s,k}^\star\bm{I}\succeq \bm{0};\label{Appen.C.18a}\\
 \left[\bm{I}-\left(\bm{X}_{e,k}^*+\mu_{e,k}^\star\bm{I}\right)\left(\bm{X}_{e,k}^*+\mu_{e,k}^\star\bm{I}\right)^{\dag}\right]\bm{X}_{e,k}^*\tilde{\bm{h}}_e=\bm{0};\label{Appen.C.18}
\end{align}
 from (\ref{p1.sdr.eqv.3}) that
\begin{align}
   &    {\rm Tr}\left\{\left(\tilde{\bm{H}}_e\bm{X}_{e,k}^*\right)\left[\bm{I} -\left(\bm{X}_{e,k}^*+\mu_{e,k}^\star\bm{I}\right)^{\dag}\bm{X}_{e,k}^*\right]\right\}+\sigma_e^2-\mu_{e,k}^{\star}\epsilon_e^2\nonumber\\
   =&\frac{\mu_{e,k} \left(P_{w_e,h_e}^\star+\sum_{i=1}^K P_{s_i,h_e}^\star\right) \lambda_{e,1}}{\mu_{e,k}+P_{w_e,h_e}^\star+\sum_{i=1}^K P_{s_i,h_e}^\star}+\sigma_e^2-\mu_{e,k}^{\star}\epsilon_e^2\nonumber\\
   \geq & \frac{\mu_{e,k} \left(\displaystyle\sum_{i\neq k}P_{s_i,h_e}^\star+P_{w_e,h_e}^\star-\frac{1}{\gamma_{e,k}}P_{s_k,h_e}^\star\right) \lambda_{e,1}}{\mu_{e,k}+\displaystyle\sum_{i\neq k}P_{s_i,h_e}^\star+P_{w_e,h_e}^\star-\frac{1}{\gamma_{e,k}}P_{s_k,h_e}^\star}+\sigma_e^2-\mu_{e,k}^{\star}\epsilon_e^2  \nonumber\\
  = & {\rm Tr}\left\{\left(\tilde{\bm{H}}_e\bm{X}_{e,k}^\star\right)\left[\bm{I} -\left(\bm{X}_{e,k}^\star+\mu_{e,k}\bm{I}\right)^{\dag}\bm{X}_{e,k}^\star\right]\right\}+\sigma_e^2-\mu_{e,k}^\star\epsilon_e^2>0;\label{Appen.C.19}
 \end{align}
 from (\ref{p1.sdr.eqv.4}) that
 \begin{align}
 \left[\bm{I}-\left(\bm{X}_{s,k}^*+\mu_{s,k}^\star\bm{I}\right)\left(\bm{X}_{s,k}^*+\mu_{s,k}^\star\bm{I}\right)^{\dag}\right]\bm{X}_{s,k}^*\tilde{\bm{h}}_k=\bm{0};\label{Appen.C.20}
  \end{align}
 from (\ref{p1.sdr.eqv.5}) that
 \begin{align}
  &    {\rm Tr}\left\{\left(\tilde{\bm{H}}_k\bm{X}_{s,k}^*\right)\left[\bm{I} -\left(\bm{X}_{s,k}^*+\mu_{s,k}^\star\bm{I}\right)^{\dag}\bm{X}_{s,k}^*\right]\right\}-\sigma_k^2-\mu_{s,k}^\star\epsilon_k^2\nonumber\\
= & \frac{\mu_{s,k}^\star \left(P_{s_k,h_k}^\star-\sum_{i\neq k}P_{s_i,h_k}^\star-P_{w_e,h_k}^\star\right)/\gamma_k \lambda_{k,1}}{\mu_{s,k}^{\star}+\left(P_{s_k,h_k}^\star-\sum_{i\neq k}P_{s_i,h_k}^\star-P_{w_e,h_k}^\star\right)/\gamma_k}-\sigma_k^2-\mu_{s,k}^\star\epsilon_k^2\nonumber\\
\geq  &  \frac{\mu_{s,k}^\star \left(P_{s_k,h_k}^\star/\gamma_k-\sum_{i\neq k}P_{s_i,h_k}^\star-P_{w_e,h_k}^\star\right) \lambda_{k,1}}{\mu_{s,k}^{\star}+P_{s_k,h_k}^\star/\gamma_k-\sum_{i\neq k}P_{s_i,h_k}^\star-P_{w_e,h_k}^\star}-\sigma_k^2-\mu_{s,k}^\star\epsilon_k^2\nonumber\\
=  & {\rm Tr}\left\{\left(\tilde{\bm{H}}_k\bm{X}_{s,k}^\star\right)\left[\bm{I} -\left(\bm{X}_{s,k}^\star+\mu_{s,k}^\star\bm{I}\right)^{\dag}\bm{X}_{s,k}^\star\right]\right\}-\sigma_k^2-\mu_{s,k}^\star\epsilon_k^2>0;\label{Appen.C.21}
 \end{align}
  from (\ref{p1.sdr.eqv.6}), (\ref{p1.sdr.eqv.7})  and (\ref{p1.sdr.eqv.8}) that
 \begin{align}
   \bm{X}_{e,k}^*=\displaystyle\sum_{i\neq k}\bm{S}_i^*+\bm{W}_e^*-\frac{1}{\gamma_{e,k}}\bm{S}_k^*, \ \bm{X}_{s,k}^*=\frac{1}{\gamma_k}\bm{S}_k^*-\sum_{i\neq k}\bm{S}_i^*-\bm{W}_e^*; \label{Appen.C.22}\\
  \mu_{e,k}^{\star}\geq 0,\quad\mu_{s,k}^{\star}\geq 0,\quad \bm{W}_e^*\succeq \bm{0}, \quad\bm{S}_k^*\succeq \bm{0}, \quad k=1,2,\ldots,K. \label{Appen.C.23}
\end{align}
Then it  can be inferred  from (\ref{Appen.C.17})$\sim$(\ref{Appen.C.23}) that
\begin{enumerate}
\item [(a)] If ${\rm Rank}\left(\bm{W}_e^\star\right)=C_e>1$, we know from (\ref{Appen.C.2}) that there exists at least one $P_{w_e,h_i}^\star>0, i\in \{1,\ldots,K\}$. Thus, from (\ref{Appen.C.17})$\sim$(\ref{Appen.C.20}), (\ref{Appen.C.22}) and (\ref{Appen.C.23}) we know that  $\bm{W}_e^*$ and $\{\bm{S}_k^*\}$ will not change the object value (\ref{p1.sdr.eqv.1})  and always satisfy constraints  (\ref{p1.sdr.eqv.2a})$\sim$(\ref{p1.sdr.eqv.4}) and (\ref{p1.sdr.eqv.6})$\sim$(\ref{p1.sdr.eqv.8}). While from (\ref{Appen.C.21}) we know that  $\bm{W}_e^*$ and $\{\bm{S}_k^*\}$ will always provide more freedoms. Thus, $\bm{W}_e^*$ and $\{\bm{S}_k^*\}$ are better solutions than $\bm{W}_e^\star$ and $\{\bm{S}_k^\star\}$.
\item [(b)] If ${\rm Rank}\left(\bm{S}_k^\star\right)=C_k>1$, we know from (\ref{Appen.C.2}) that there exists at least one $P_{s_k,h_{i\neq k}}^\star>0, i\in \{1,\ldots,K\}$ or $P_{s_k,h_{e}}^\star>0$. Then, if $P_{s_k,h_{i\neq k}}^\star>0, i\in \{1,\ldots,K\}$, we know from (\ref{Appen.C.17}) that $\bm{W}_e^*$ and $\{\bm{S}_k^*\}$ will always provide smaller objective value than $\bm{W}_e^\star$ and $\{\bm{S}_k^\star\}$. While from (\ref{Appen.C.18a})$\sim$(\ref{Appen.C.23}), we know that $\bm{W}_e^*$ and $\{\bm{S}_k^*\}$ satisfy all the constraints of $\mathbf{P1\!-\!SDR\!-\!EQV}$. Thus, $\bm{W}_e^*$ and $\{\bm{S}_k^*\}$ are better solutions. If $P_{s_k,h_{e}}^\star>0$, we know from (\ref{Appen.C.17})$\sim$(\ref{Appen.C.18})  and (\ref{Appen.C.20})$\sim$(\ref{Appen.C.23}) that $\bm{W}_e^*$ and $\{\bm{S}_k^*\}$ will not change the object value (\ref{p1.sdr.eqv.1})   and always satisfy constraints (\ref{p1.sdr.eqv.2a}), (\ref{p1.sdr.eqv.2}) and (\ref{p1.sdr.eqv.4})$\sim$(\ref{p1.sdr.eqv.8}). While from (\ref{Appen.C.19}) we conclude that  $\bm{W}_e^*$ and $\{\bm{S}_k^*\}$ will always provide more freedoms. Thus, $\bm{W}_e^*$ and $\{\bm{S}_k^*\}$ are better solutions than $\bm{W}_e^\star$ and $\{\bm{S}_k^\star\}$.
\end{enumerate}
Based on the above discussions, we know that $\bm{W}_e^*$ and $\{\bm{S}_k^*\}$ are always better solutions, which contradicts the assumption that $\bm{W}_e^\star$ and $\{\bm{S}_k^\star\}$ are optimal solutions. Thus there must be
${\rm Rank}\left(\bm{W}_e^\star\right)\leq 1$ and ${\rm Rank}\left(\bm{S}_k^\star\right)\leq 1, \ \forall k\in \{1,\ldots,K\}$.
From (\ref{Appen.C.12}), we know $\bm{W}_e^\star$ and $\{\bm{S}_k^\star\}$ can be further expressed as
\begin{align}
\bm{W}_e^*=\displaystyle P_{w_e}^\star\frac{\tilde{\bm{h}}_e\tilde{\bm{h}}_e^{\rm H}}{\|\tilde{\bm{h}}_e\|^2},\quad
\bm{S}_k^*=\displaystyle P_{s_k}^\star\frac{\tilde{\bm{h}}_k\tilde{\bm{h}}_k^{\rm H}}{\|\tilde{\bm{h}}_k\|^2},
\end{align}
where $P_{w_e}^\star$ and $P_{s_k}^\star$ are the optimal power allocation for AN beamforming and information beamforming, respectively.
The proof of Theorem~1 is thus completed.


\end{document}